\documentclass[twocolumn]{aastex631}
\pdfoutput=1 
 \usepackage{amsmath,amstext}
\usepackage[T1]{fontenc}
\usepackage{apjfonts} 
\usepackage[figure,figure*]{hypcap}
\usepackage{url}
\usepackage{CJKutf8}

\shorttitle{Critical Metallicity and Mass Loss}
\shortauthors{Ou et al.}

\begin{document}
\begin{CJK*}{UTF8}{bsmi}
\title{Critical Metallicity of Cool Supergiant Formation. I. Effects on Stellar Mass Loss and Feedback}

\author[0000-0003-1295-8235]{Po-Sheng Ou （歐柏昇）}
\affiliation{Institute of Astronomy and Astrophysics, Academia Sinica, No.1, Sec. 4, Roosevelt Rd., Taipei 10617, Taiwan, R.O.C.} 
\affiliation{Department of Physics, National Taiwan University, No.1, Sec. 4, Roosevelt Rd.,  Taipei 10617, Taiwan, R.O.C.}

\correspondingauthor{Po-Sheng Ou}
\email{psou@asiaa.sinica.edu.tw}

\author[0000-0002-4848-5508]{Ke-Jung Chen（陳科榮）}
\affiliation{Institute of Astronomy and Astrophysics, Academia Sinica, No.1, Sec. 4, Roosevelt Rd., Taipei 10617, Taiwan, R.O.C.} 
	
\author[0000-0003-3667-574X]{You-Hua Chu （朱有花）}
\affiliation{Institute of Astronomy and Astrophysics, Academia Sinica, No.1, Sec. 4, Roosevelt Rd., Taipei 10617, Taiwan, R.O.C.} 
\affiliation{Department of Physics, National Taiwan University, No.1, Sec. 4, Roosevelt Rd.,  Taipei 10617, Taiwan, R.O.C.} 
\affiliation{Department of Astronomy, University of Illinois, 1002 West Green Street, Urbana, IL 61801, U.S.A. }

\author[0000-0001-5466-8274]{Sung-Han Tsai（蔡松翰）}
\affiliation{Institute of Astronomy and Astrophysics, Academia Sinica, No.1, Sec. 4, Roosevelt Rd., Taipei 10617, Taiwan, R.O.C.} 
\affiliation{Department of Physics, National Taiwan University, No.1, Sec. 4, Roosevelt Rd.,  Taipei 10617, Taiwan, R.O.C.} 

\begin{abstract}
This paper systematically studies the relation between metallicity and mass loss of massive stars. We perform one-dimensional stellar evolution simulations and build a grid of $\sim$2000 models with initial masses ranging between 11 and 60 $M_{\odot}$ and absolute metallicities $Z$ between 0.00001 and 0.02. Steady-state winds, comprising hot main-sequence winds and cool supergiant winds, are the main drivers of the mass loss of massive stars in our models. We calculate the total mass loss over the stellar lifetime for each model. Our results reveal the existence of a critical metallicity $Z_{\rm{c}}$ at $Z \sim 10^{-3}$, where the mass loss exhibits a dramatic jump. If $Z>Z_{\rm{c}}$, massive stars tend to evolve into cool supergiants, and a robust cool wind is operational. In contrast, if $Z<Z_{\rm{c}}$, massive stars usually remain as blue supergiants, wherein the cool wind is not activated and the mass loss is generally weak. Moreover, we calculate the wind feedback in a $10^5$ $M_{\odot}$ star cluster with the Salpeter initial mass function. The kinetic energy released by winds does not exhibit any significant transition at $Z_{\rm{c}}$ because the wind velocity of a cool supergiant wind is low and contributes little to the kinetic energy. The effects of critical metallicity provide implications for the fates of metal-poor stars in the early universe.

\end{abstract}


\section{Introduction}\label{sec:intro}

Massive stars lose a substantial amount of mass via stellar winds, which inject energies and enriched material  into the interstellar medium (ISM). Main-sequence OB stars can produce \textit{hot winds} that are driven by the radiation pressure of the line emission \citep[][for review]{puls2008,vink2021}, and red supergiants (RSGs) produce even stronger \textit{cool winds} that are driven by the dust continuum \citep[][for review]{willson2000,decin2021}. The mass loss from these winds influences the evolutionary tracks and the fates of massive stars \citep{langer2012,smith2014,vink2015} and reshapes the ambient environment \citep[e.g.,][]{chu2008}.

The mass from the stripped-off envelopes of massive stars form circumstellar media (CSM). 
When a massive star explodes as a supernova (SN), it may interact with the CSM produced by its progenitor star. This interaction plays an important role in the SN observables \citep[e.g.,][]{smith2014,maeda2015,moriya2017,chen2020}. 
If the SN explosion was proceeded by a blue supergiant (BSG) or Wolf-Rayet (WR) phase, its fast wind would have swept up the CSM to form a wind-blown bubble. The SN ejecta and shock wave will encounter the circumstellar bubble, and strongly affect the dynamical evolution of SNR \citep[e.g.,][]{dwarkadas2005,patnaude2017,ou2018}. Therefore, stellar winds have a profound impact on SNe and their resulting SNRs, which drive the ISM ecosystem.

On a larger scale, stellar winds serve as a critical ingredient of stellar feedback in galaxies \citep{abbott1982,leitherer1992,freyer2003}. Although the contribution of  stellar winds is thought to be less powerful than that of SNe \citep[e.g.,][]{walch2015,dale2015}, some ISM simulations indicate that stellar winds are more efficient than SNe in disrupting giant molecular clouds \citep{rogers2013,fierlinger2016,reyraposo2017} and retaining energies in superbubbles \citep{krause2013}. For cosmological simulations, stellar feedback is essential for reproducing the observed star-formation rates in galaxies, and appropriate expressions of stellar mass loss are often required \citep[e.g.,][]{stinson2006,hopkins2014,agertz2015,schaye2015,wang2015,hopkins2018,pillepich2018, springel2018,emerick2019,marinacci2019,hopkins2022}.

To evaluate the energy feedback of massive stars, we need to consider the total mass loss and total kinetic energy injected by stellar winds. The mass loss rate ($\dot{M}$) of a stellar wind depends on metallicity ($Z$), and the dependence is typically expressed as $\dot{M} \sim Z^m$, where $m$ is a constant \citep[e.g.,][]{vink2001,mokiem2007,mauron2011}. In the aforementioned cosmological simulations, a single-value $Z$ appropriate for a particular epoch is used for the $\dot{M}$ and energy feedback estimates. 
It is well known that $Z$ can strongly impact the evolutionary tracks of stars \citep{eleid2009,maeder2009,langer2012}, and $Z$ varies along the stellar evolution tracks. It is conceivable that the dependence of $\dot{M}$ on $Z$ varies along the stellar evolution tracks.  To estimate the total mass loss ($\Delta M$) of a massive star, it is necessary to integrate the $\dot{M}$ for its contemporary $Z$ over the entire evolutionary track.  Likewise, the total kinetic energy injected by a massive star's stellar winds has to be determined by integrating the wind mechanical luminosity over the evolutionary track using $\dot{M}$ and wind velocity appropriate for their contemporary $Z$.

Previous calculations of $\Delta M$ of massive stars have been carried out for a wide range of initial masses \citep[e.g.,][]{limongi2017,renzo2017,beasor2018}. It is found that different wind prescriptions cause $\sim$50\% uncertainties in the resulting $\Delta M$ \citep{renzo2017}. Recently, \citet{fichtner2022} calculated the total mass and energy released by stellar winds within a star cluster and showed that they generally increase with metallicity; they also found that the rate of increase is quite different for single and binary stars. The above studies either considered an entire cluster or included only a small number of discrete $Z$ values; thus they do not allow a detailed examination of metallicity effects.  

We have systematically studied the metallicity effects on the total mass loss and kinetic energy release of massive stars by employing the MESA code to perform a series of stellar evolution simulations. We establish models of single non-rotating stars with wind physics. The numerical methods and stellar wind prescriptions are described in Section \ref{sec:method}. Based on this model grid, we describe the general features of stellar evolution in Section \ref{sec:evolution}. Then in Section \ref{sec:massloss}, we present the main results of the total mass loss as a function of the initial mass and metallicity of massive stars. We report a \textit{critical metallicity} ($Z_{\rm c}$) of mass loss and show that it is associated with the cool supergiant formation. In Section \ref{sec:feedback} we calculate the energy feedback of winds using our grid of models. Finally, we discuss the implications of our results in Section \ref{sec:discussions} and present the conclusions in Section \ref{sec:conclusions}. In the companion Paper II, we will further investigate the physical mechanism of $Z_{\rm c}$ and explain it according to the fundamental understanding of cool supergiant formation.

\section{Methodology}\label{sec:method}
In this section, we introduce the stellar evolution code for building our models, describe the prescriptions of stellar mass loss, and summarize the parameter space of our models.

\subsection{MESA Code}
We perform massive-star evolution simulations using the Modules for Experiments in Stellar Astrophysics  \citep[MESA;][]{paxton2011,paxton2013,paxton2015,paxton2018,paxton2019} version No.10108. We primarily use \texttt{MESAstar}, a 1D stellar evolution code that can model stellar evolution from the pre-main-sequence stage to the zero-age main sequence (ZAMS) stage, and from the ZAMS to the iron core-collapse stage. It hydrodynamically evolves stars using microphysics such as nuclear reaction networks, equations of state, and opacities, which are the key physics of stellar evolution and are available in MESA. 

MESA provides several stellar wind functions that can be used to calculate the mass-loss rates using stellar parameters at each timestep during the stellar evolution. When a mass-loss routine is turned on, the code adjusts the stellar structure by rescaling the mass coordinates of the mass cells, according to the mass-loss rate given by the wind functions \citep{paxton2011}.

\subsection{Physical Processes}

Our simulations include two steps. First we evolve a pre-main-sequence star to reach the ZAMS stage, and then, we use the ZAMS model as the initial condition in the stellar evolution and evolve the model from the main sequence to the end of the star's life. Different nuclear reaction networks are employed in these two steps. For the pre-main-sequence evolution, we simply adopt the “basic” network with eight isotopes. For the evolution from the ZAMS stage onward, we employ the more complicated nuclear reaction network “approx21 (Cr60),” which includes 21 isotopes \footnote{$^1$H, $^3$He, $^4$He, $^{12}$C, $^{14}$N, $^{16}$O, $^{20}$Ne, $^{24}$Mg, $^{28}$Si, $^{32}$S, $^{36}$Ar, $^{40}$Ca, $^{44}$Ti, $^{48}$Cr, $^{52}$Fe, $^{54}$Fe, $^{56}$Fe, $^{56}$Ni, $^{60}$Cr, proton, neutron}.

In the stellar evolution simulations, we mainly use the default parameters for massive stars in the “inlist\_massive\_defaults” file in the \texttt{MESAstar} module. Additionally, the Ledoux criterion and Henyey's mixing-length theory \citep{henyey1965} are applied, and the mixing-length scaling parameter $\alpha$ (i.e., the ratio of mixing length to pressure scale height) is set at 1.5. 
The Type 2 OPAL opacity tables \citep{iglesias1996} are applied during and after the helium burning. In addition to the default parameters, we limit the change of helium abundance
per timestep to $\leq$10$\%$ to prevent abrupt changes.

\subsection{Mass-Loss Prescriptions}

Although the mass-loss rates can be calculated using theoretical approaches \citep[e.g.,][]{vink2000,vink2001}, such calculations are time-consuming and only available for hot winds. Instead, simple formulae of the mass-loss rates derived from observations or theoretical calculations are usually adopted in stellar evolution. Most of these widely used mass-loss prescriptions have been built in as modules in MESA. In our stellar evolution simulations, we apply the hot wind prescription of \citet[][hereinafter, V01]{vink2000,vink2001}, cool wind prescription of \citet[][hereinafter, dJ88]{dejager1988}, and WR wind prescription of \citet[][hereinafter, NL00]{nugis2000}. These prescriptions are described below.

For hot OB stars, metal ions in the stellar atmospheres absorb photons in the ultraviolet (UV) resonance lines and are radiatively accelerated, yielding steady-state winds
 \citep{lucy1970,castor1975,puls2008}. 
Using multi-line radiative transfers calculated by the Monte Carlo methods  \citep{abbott1985}, \citet{vink2000,vink2001} presented a grid of wind models, fitted the models with a linear regression, and established a mass-loss prescription for hot stars with effective temperatures $T_{\textrm{eff}}>10^4$ K.

When a massive star evolves into an RSG, its cool wind is driven by the radiation pressure on dust grains, similar to the winds from the asymptotic-giant-branch stars. To date, although empirical laws have been established, no well-established theoretical predictions for cool winds exist, because the origin and structure of RSG winds are very complex and the cool wind mechanism is poorly understood \citep{smith2014}. 
Among the mass-loss rates suggested by the various empirical laws, up to a factor of 10 scatter has been seen \citep[dJ88;][]{reimers1975,nieuwenhuijzen1990,vanloon2005,mauron2011,goldman2017,beasor2018,beasor2020,humphreys2020}. In our simulations, we adopt the most commonly used cool-wind prescription by dJ88, which is an empirical function based on the fitting of the observed mass-loss rates of a set of O–M stars.
Although $\dot{M}\sim Z^m$ has been included in some recent cool wind functions \citep[e.g.,][]{mauron2011,groh2019}, the exponent $m$ is uncertain. 
Recent observations have shown that the mass-loss rates of RSGs are nearly independent of metallicity \citep{goldman2017,beasor2020}. Therefore, we adopt the dJ88 prescription without metallicity dependence available in MESA, similar to other works such as \citet{choi2016}.

WR stars have $T_{\textrm{eff}}>10^4$ K, but their mass-loss rates are a factor of 10 higher than those of hot O stars exhibiting the same luminosity \citep{nugis2002}; thus, the hot-wind prescription is not suitable for them. We adopt the WR wind prescription by NL00, which is an empirical law of mass-loss rate as a function of luminosity, helium abundance, and metallicity. It was derived from the observations of 64 Galactic WR stars with the correction of clumping in the wind. This prescription is supported by an optically thick radiation-driven wind theory \citep{nugis2002}. 

The hot, cool, and WR wind schemes operate under different stellar evolution stages. The switches between different wind schemes depend on $T_{\textrm{eff}}$ and the surface hydrogen abundance ($X_H$) of the simulated stars. The wind prescriptions are explained in detail in Appendix \ref{sec:appA}.

In the “Dutch” hot/WR wind scheme in MESA, a “low-T” regime is present where the dJ88 prescription is adopted rather than the V01 prescription. For simplicity, we use the term “hot wind” to denote only the mass loss derived from the V01 prescription, and use the term “cool wind” to denote all the mass loss derived from the dJ88 prescription.

\subsection{Stellar Evolution Simulations}

We establish a set of models to investigate the influence the initial mass and metallicity on the stellar mass loss. In our grid, the initial mass values are sampled successively in the range of 11–60 $M_{\odot}$ with 1-$M_{\odot}$ steps. Moreover, 38 different values of initial absolute metallicities $Z$ are chosen: $Z=(1,2,...,9) \times 10^{-5}$, $(1,2,...,9) \times 10^{-4}$, $(1,2,...,9) \times 10^{-3}$, and $(1.0,1.1,1.2,...,2.0) \times 10^{-2}$. For reference, the solar metallicity is $Z_{\odot} = 1.34\times 10^{-2}$ \citep{asplund2009}. Overall, the total number of models in our grid is 1900. 

Our stellar models are typically evolved until the iron core-collapse stage. However, some models with initial masses of 11–17 $M_{\odot}$ stall at the carbon or oxygen burning stages, usually because of abrupt changes in abundance that cannot be easily treated numerically. As these stars have already passed the RSG phase,
we adopt the masses at this point as the final masses.

We perform our simulations using the TIARA-TC cluster \footnote{\url{https://www.tiara.sinica.edu.tw/facility.php}} maintained by the Institute of Astronomy and Astrophysics, Academia Sinica (ASIAA). This cluster has 50 nodes, 600 CPU cores, and 2.3 TB memory. Our grid simulations were completed in about 100,000 core-hours.

\section{Stellar Evolution and Mass Loss Rate}\label{sec:evolution}

Stellar mass loss depends on the physical parameters of stars, which vary along the stellar evolution. In this section, we first describe the stellar evolution and then their associated mass loss.

\subsection{Stellar Evolution}
We preform simulations of stellar evolution with different initial masses and metallicities. Their evolutionary tracks in the theoretical Hertzprung–Russell (HR) diagram, i.e., luminosity ($L$) versus $T_{\textrm{eff}}$ diagram, are shown in the left panels of Figure 1. During the main-sequence phase of $20-60\,M_{\odot}$ stars, $T_{\textrm{eff}}$ decreases by 0.1--0.25 dex and $L$ increases by 0.3--0.15 dex over the main-sequence lifetime of $\sim$ 8--4 Myr. 
A star reaches the terminal-age main sequence (TAMS) when the hydrogen in its stellar core is exhausted. 
At this time, a hydrogen-burning shell forms, the star starts expanding, and $T_{\textrm{eff}}$ quickly decreases. 
Then, the core-helium burning starts, and the outer envelope continues to expand and becomes much cooler.
In the HR diagram, the star quickly crosses the “Hertzsprung gap” and moves to the right side. 
After these post-main-sequence evolution stages, the star becomes a supergiant. 
Supergiants with $\log (T_{\textrm{eff}}/\textrm{K})<3.6$ (or $T_{\textrm{eff}}\lesssim$ 4,000 K) are often called RSGs,
$3.6<\log (T_{\textrm{eff}}/\textrm{K})<3.8$ (or 4,000 K $\lesssim T_{\textrm{eff}}\lesssim$ 6,300 K) are called yellow supergiant (YSGs), and $\log (T_{\textrm{eff}}/\textrm{K})>3.8$ (or $T_{\textrm{eff}}\gtrsim$ 6,300 K) are called BSGs \citep[e.g.,][]{georgy2012}. 

A star's lowest $T_{\textrm{eff}}$ in its evolutionary track depends on its initial mass and metallicity. As shown in Figure~\ref{fig:HR+mdot}, for stars with $Z=0.02$, those with initial masses $\gtrsim$50 $M_{\odot}$ can only become YSGs instead of RSGs. Furthermore, comparison of the models of $Z=0.02$ and $Z=0.0005$ shows that the evolutionary tracks of lower-$Z$ stars usually end at higher $T_{\textrm{eff}}$. This feature is critical in the relation between mass loss and metallicity, and will be elaborated in later sections.
For convenience, we use the term “cool supergiant” as long as the dJ88 wind prescription is fully active, i.e. $T_{\textrm{eff}}<$ 10,000 K, or $\log (T_{\textrm{eff}}/\textrm{K})<4.0$. 

During the advanced burning stages, stars with initial masses $\gtrsim 50$ $M_{\odot}$ can reach luminosities $\gtrsim 10^6$ $L_{\odot}$. These stars can be regarded as luminous blue variables (LBVs)  \citep{humphreys1994}, but our simulations do not include eruptive mass loss, which is a common feature of LBVs. Furthermore, some LBVs lose their envelopes and evolve back to a hotter region in the HR diagram to become WR stars. We use the criterion of surface hydrogen abundance ($X_S$) below 0.4 to identify WR stars in our work.

\subsection{Evolution of Mass Loss Rate}\label{sec:mdot_evolution}

\begin{figure*}[tbh]
\centering
\includegraphics[width=\columnwidth]{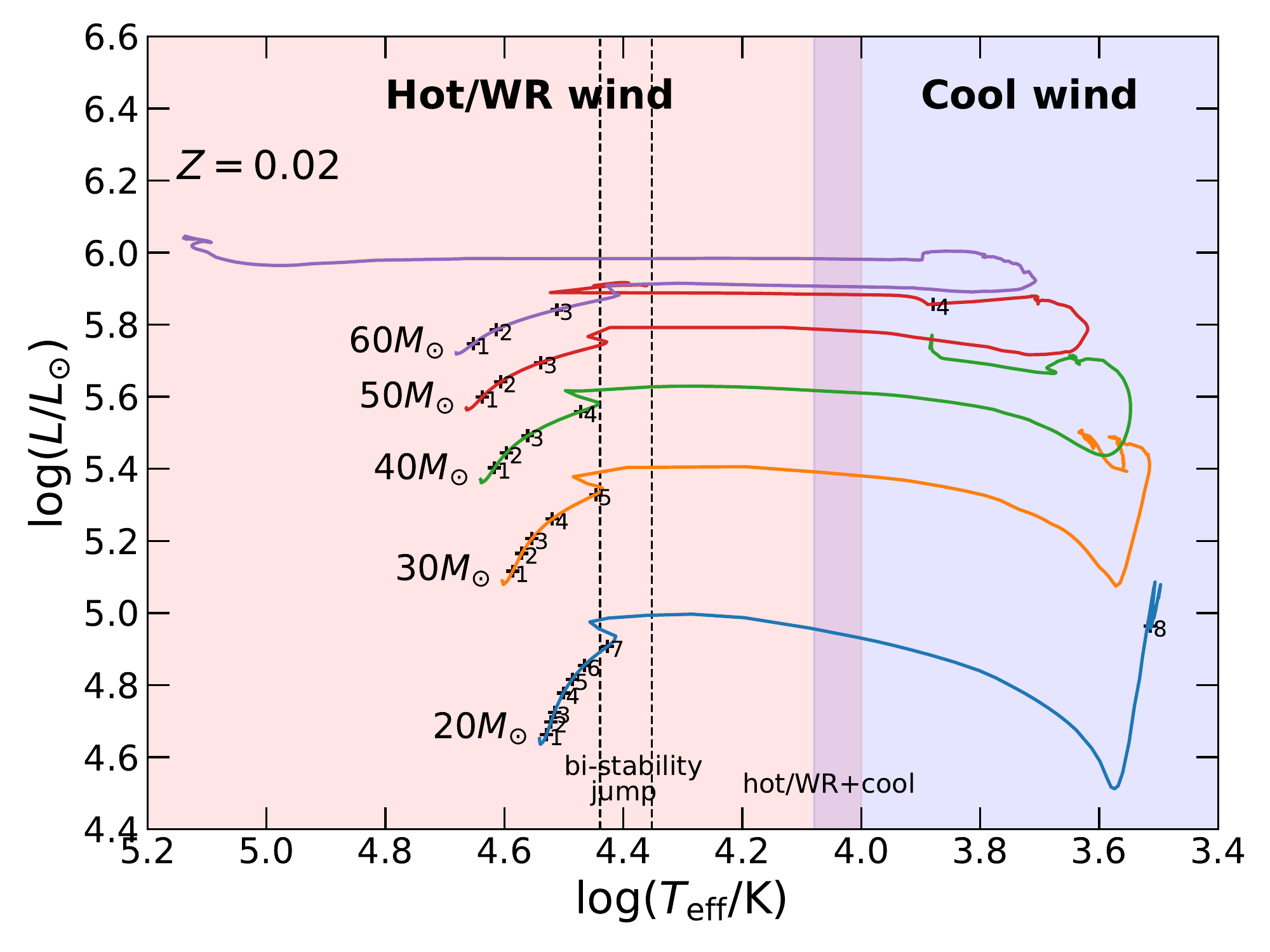}
\includegraphics[width=\columnwidth]{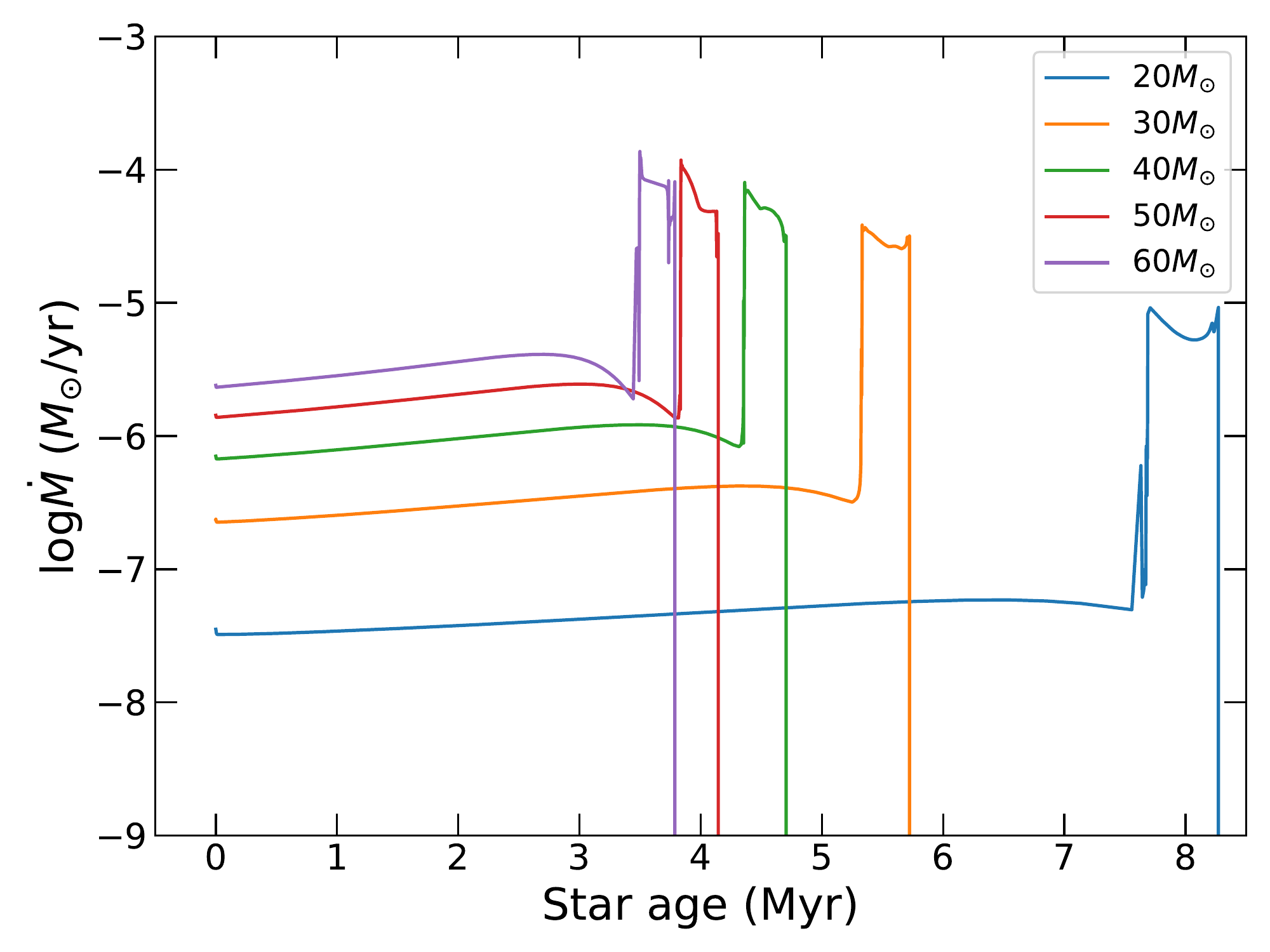}
\includegraphics[width=\columnwidth]{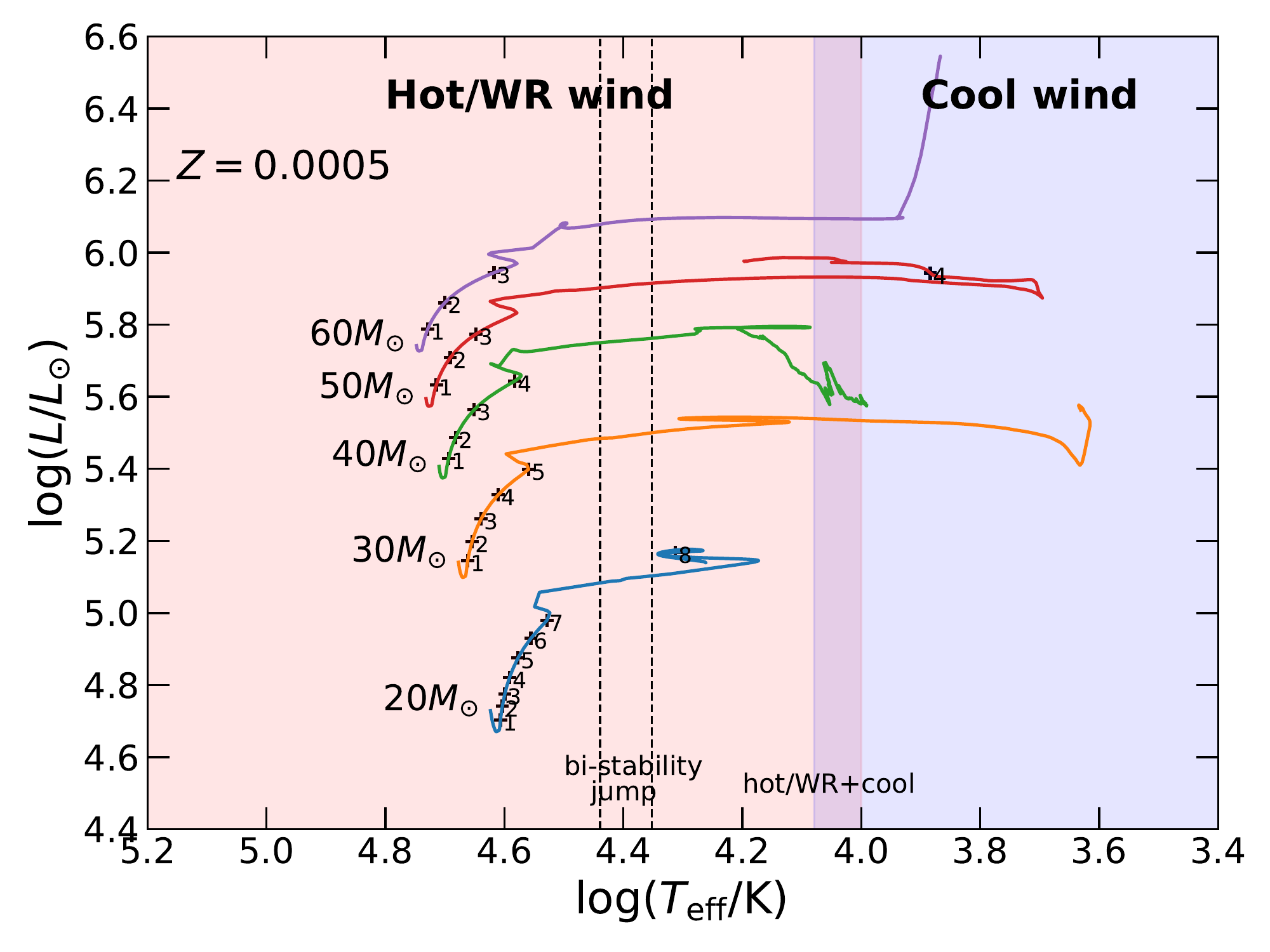}
\includegraphics[width=\columnwidth]{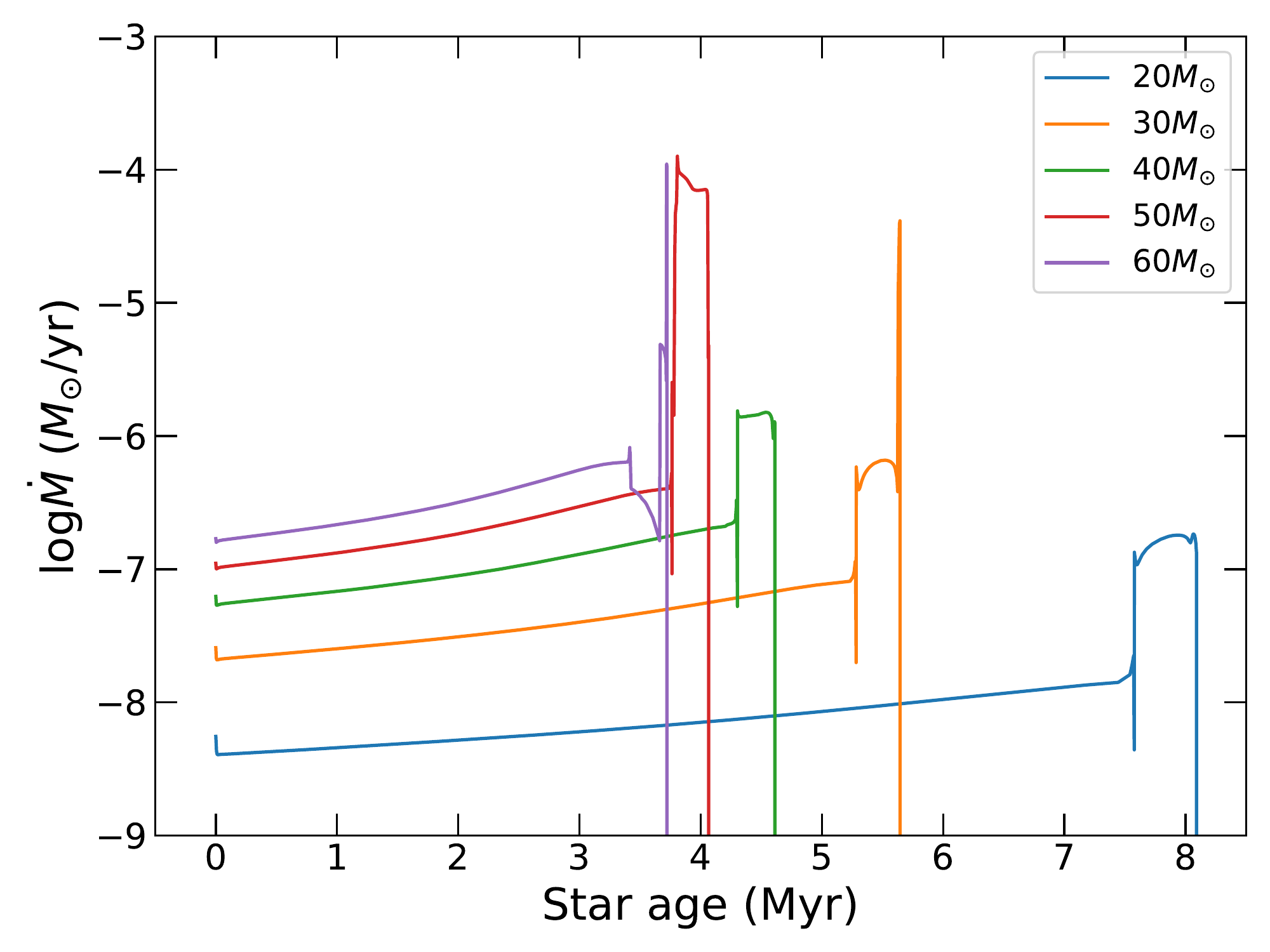}
\caption{The HR diagram of massive stars and their mass-loss histories. \textit{Left panels}: the stellar evolutionary tracks of $20-60$ $M_{\odot}$ stars with $Z =0.02$ and 0.0005. 
 The region between dashed lines denotes the temperature range of the bi-stability jump. 
 The red and blue regions represent hot and cool winds, respectively, instead of the colors of stars. When 10,000 K$<T_{\rm{eff}}<$12,000 K, a combination of hot and cool winds is adopted. The "+" marks and the numbers beside them record the stellar ages in the unit of Myr.
\textit{Right panels}: The evolution of the mass-loss rates of stars with absolute metallicities $Z=2\times 10^{-2}$ (\textit{top panel}) and $Z=5\times 10^{-4}$ (\textit{bottom panel}). The mass loss of massive stars depends on their initial masses and metallicities, and their mass-loss rates in the post-main sequence can be $10^{3} - 10^{4}$ higher than those in the main sequence.}
\label{fig:HR+mdot}
\end{figure*}

As a star evolves through different burning phases, its $T_{\textrm{eff}}$ and radius vary,
resulting in different types of stellar winds and mass-loss rates. As described in Section 2.3, the mass-loss rates are calculated using a combination of mass-loss prescriptions chosen according to $T_{\textrm{eff}}$ and $X_S$. In the HR diagrams in Figure~\ref{fig:HR+mdot}, most high-$Z$ and some low-$Z$ stars shift from the hot wind region to the cool wind region in the post-main-sequence stage. The massive stars of 50 and 60 $M_{\odot}$ with $Z=0.02$ further evolve back to the hot region and enter the WR phase. The criterion $X_S<0.4$ is used to activate the WR wind scheme.

The right panels of Figure~\ref{fig:HR+mdot} present the evolution of the mass-loss rates in our stellar evolution simulations. During the main-sequence phase, $\log T_{\textrm{eff}}$ is in the range of $\sim 4.45-4.75$, and thus, hot winds dominate. For models of the same $Z$, higher-mass stars have higher luminosities and higher mass-loss rates; furthermore, a main-sequence star's mass-loss rate is usually low and stable because the $T_{\textrm{eff}}$ and luminosity changes are not large. For models of different $Z$ but the same initial mass, a higher $Z$ leads to a higher mass-loss rate, but the $\dot{M}$ dependence on $Z$ is not a simple $Z^m$ because $\dot{M}$ also depends on $L$ and $T_{\textrm{eff}}$, which are also dependent on $Z$. 
For $Z=0.0005$, $\dot{M}$ for 20--60 $M_{\odot}$ main-sequence stars is in the range $\sim 10^{-8}-10^{-6}$ $M_{\odot}$ yr$^{-1}$, and for $Z=0.02$, $\sim 10^{-7.5}-10^{-5.5}$ $M_{\odot}$ yr$^{-1}$. 
When a star evolves into the post-main-sequence stage, $T_{\textrm{eff}}$ suddenly decreases. The change in $T_{\textrm{eff}}$ causes the enhancement of mass-loss rate by about 1–3 orders of magnitude, because the $T_{\textrm{eff}}$ passes through two jumps in the mass-loss rate included in the wind prescriptions. These two jumps correspond to two physical transitions in the supergiant winds:

(1) \textit{The “bi-stability jump” in early-type supergiant winds.} When the $T_{\textrm{eff}}$ is less than $\sim$25,000 K, the mass-loss rate suddenly increases and the wind velocity decreases, which is called the “bi-stability jump” \citep{pauldrach1990}. This is caused by the recombination of Fe IV. When the temperature decreases below $\sim$25,000 K, the ionization fraction of Fe III below the sonic point significantly increases, and Fe III lines are more efficient drivers of winds than Fe IV lines \citep{vink1999}. The mass-loss prescription of V01 illustrates the bi-stability jump by setting different mass-loss functions in the two regimes with $T_{\textrm{eff}}$ values higher and lower than the jump temperature, which ranges between 22,500 and 27,500 K. 

(2) \textit{The transition from hot winds to cool supergiant winds.} When the $T_{\textrm{eff}}$ decreases below 10,000--12,000 K, the star evolves into a cool supergiant. The cool, dust-driven wind becomes the dominating means of mass loss, and the mass-loss rate significantly increases. 

These two jumps in the mass-loss rate occur at the post-main-sequence stage, marked as the bi-stability jump and the hot-to-cool wind transition region in Figure~\ref{fig:HR+mdot}.  The stellar evolution timescale between these two jumps are usually so short, $<10^4$ yr, that these two jumps are unresolvable in the evolution of mass loss rate plots (the right panels of Figure~\ref{fig:HR+mdot}).

Nevertheless, for some low-$Z$ stars with $Z=0.0005$, their $T_{\textrm{eff}}$ only decreases to the range between 10,000 and 25,000 K at the onset of the post-main-sequence stage, and they remain as early-type supergiants. Some of the other low-$Z$ stars do cool down to $T_{\textrm{eff}}<10,000$ K, but this happens in the very late evolutionary stage after the core-carbon burning commences; thus, the separation between the two jumps is about several hundred thousand years. For example, for the star with 30 $M_{\odot}$ and $Z$=0.0005, the second jump in the mass-loss rate occurs at the very end of its lifetime.

\section{Total Mass Loss and Metallicity}\label{sec:massloss}
Since stellar mass loss is very sensitive to the evolutionary phases, the total mass loss in the lifetime of a star depends on its initial mass and metallicity. This section presents our results of the total mass loss integrated through stellar lifetimes.

\subsection{Initial Mass and Metallicity Dependence of Mass Loss}

\begin{figure}[tbh]
\centering
\includegraphics[width=\columnwidth]{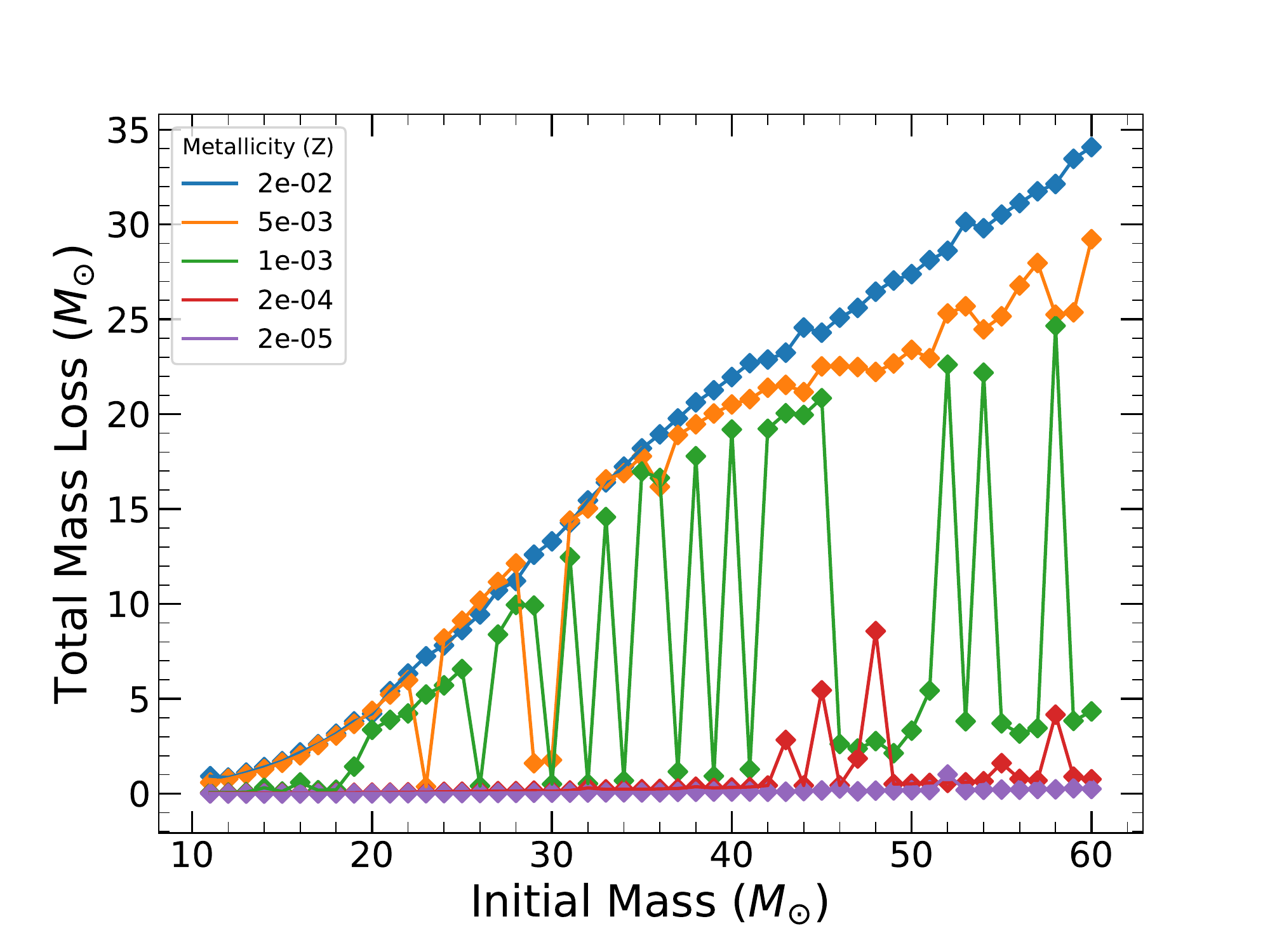}
\caption{Total mass loss as a function of the initial mass for stars with different initial metallicities.}
\label{fig:mtot_m}
\end{figure}

Figure~\ref{fig:mtot_m} displays representative relations between a star's initial mass ($M_i$) and the total mass loss ($\Delta M$), which is the integrated mass loss over the stellar lifetime. 
The total mass loss generally increases with the initial stellar mass, but the detailed relation depends on the metallicity.
At a first glance, three different trends of mass loss are observed at different metallicities:

(1) For  $Z \gtrsim 5\times 10^{-3}$, the high-$Z$ cases, the total mass loss of a star generally increases with its initial mass. For example, for $Z=0.02$ ($>1\,Z_{\odot}$ = 0.0134), a $15\,M_{\odot}$ star loses $\sim 2\,M_{\odot}$, and a $60\,M_{\odot}$ star loses $>30\,M_{\odot}$ during its lifetime. 

(2) For $Z \lesssim 3\times 10^{-4}$, the low-$Z$ cases, the total mass loss of a star also increases with its initial mass, but it is considerably lower than that of the $Z \gtrsim 10^{-2}$ cases. For example, for $Z=10^{-4}$, a 15 $M_{\odot}$ star loses $\sim 0.01$ $M_{\odot}$, and a $60\,M_{\odot}$ star loses $\sim 0.5\,M_{\odot}$. Some low-$Z$ high-mass stars have high mass losses, but their total mass loss is still $<20 \%$ of the initial mass.

(3) For stars with $Z\sim 10^{-3}$, the total mass loss is not a smooth function of the initial mass; instead, it oscillates between the high values for the high-$Z$ cases and the low values for the low-$Z$ cases.

\begin{figure}[tbh]
\centering
\includegraphics[width=\columnwidth]{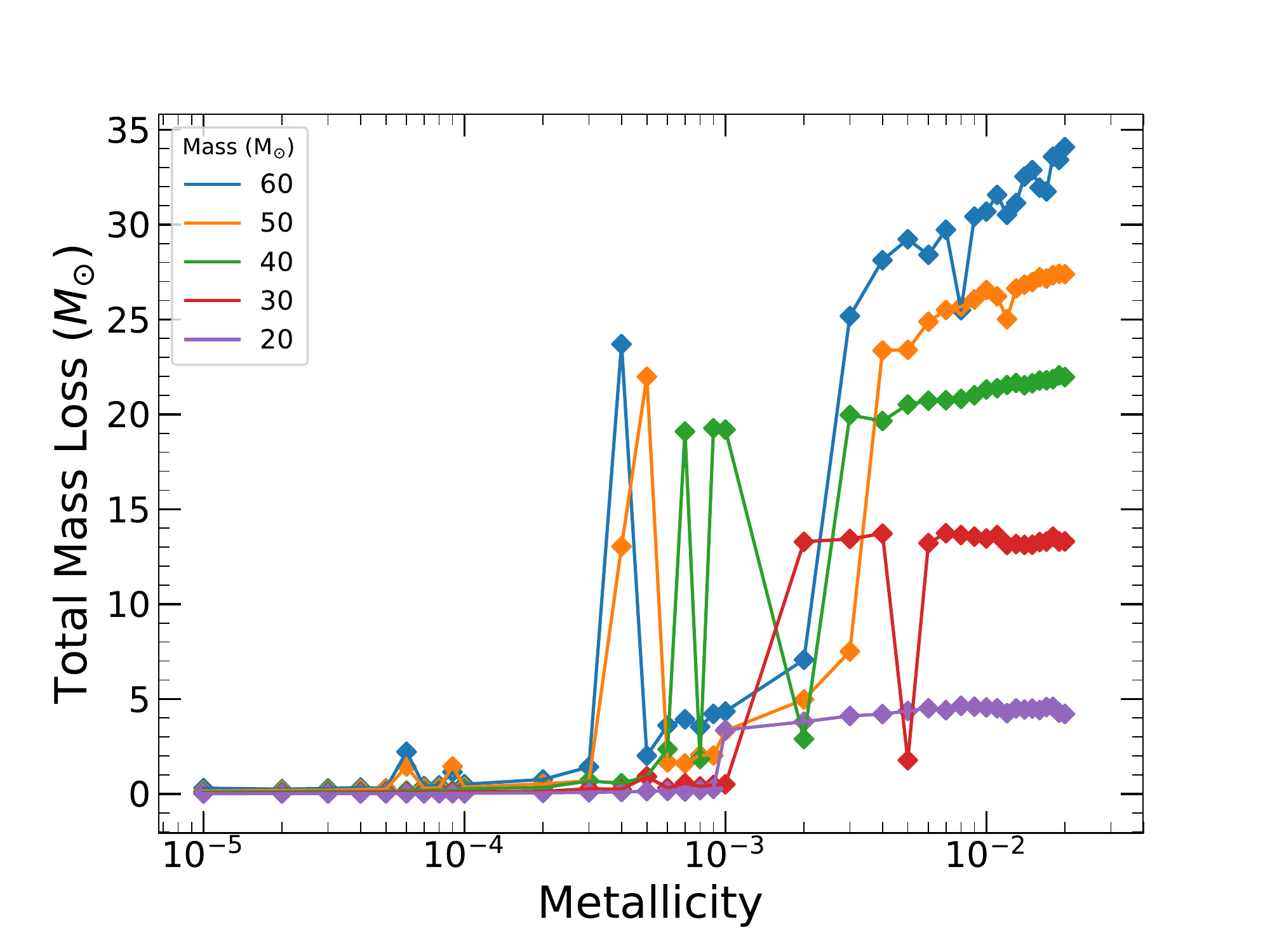}
\caption{Total mass loss as a function of initial metallicity for stars with different initial masses.}
\label{fig:mtot_Z}
\end{figure}

\subsection{Critical Metallicity}

We have shown that the total mass loss during a star's lifetime largely depends on its metallicity. To further illustrate the metallicity effects on mass loss, we plot the total mass loss as a function of metallicity in Figure~\ref{fig:mtot_Z}. 

Figure~\ref{fig:mtot_Z} displays a significant jump in the total mass loss at $Z\sim 10^{-3}$. When $Z \lesssim 3\times 10^{-4}$, the mass loss is generally low. When $Z$ approaches $\sim 10^{-3}$, the mass loss jumps to a significantly higher level. 

\begin{figure*}[tbh]
\centering
\includegraphics[width=15cm]{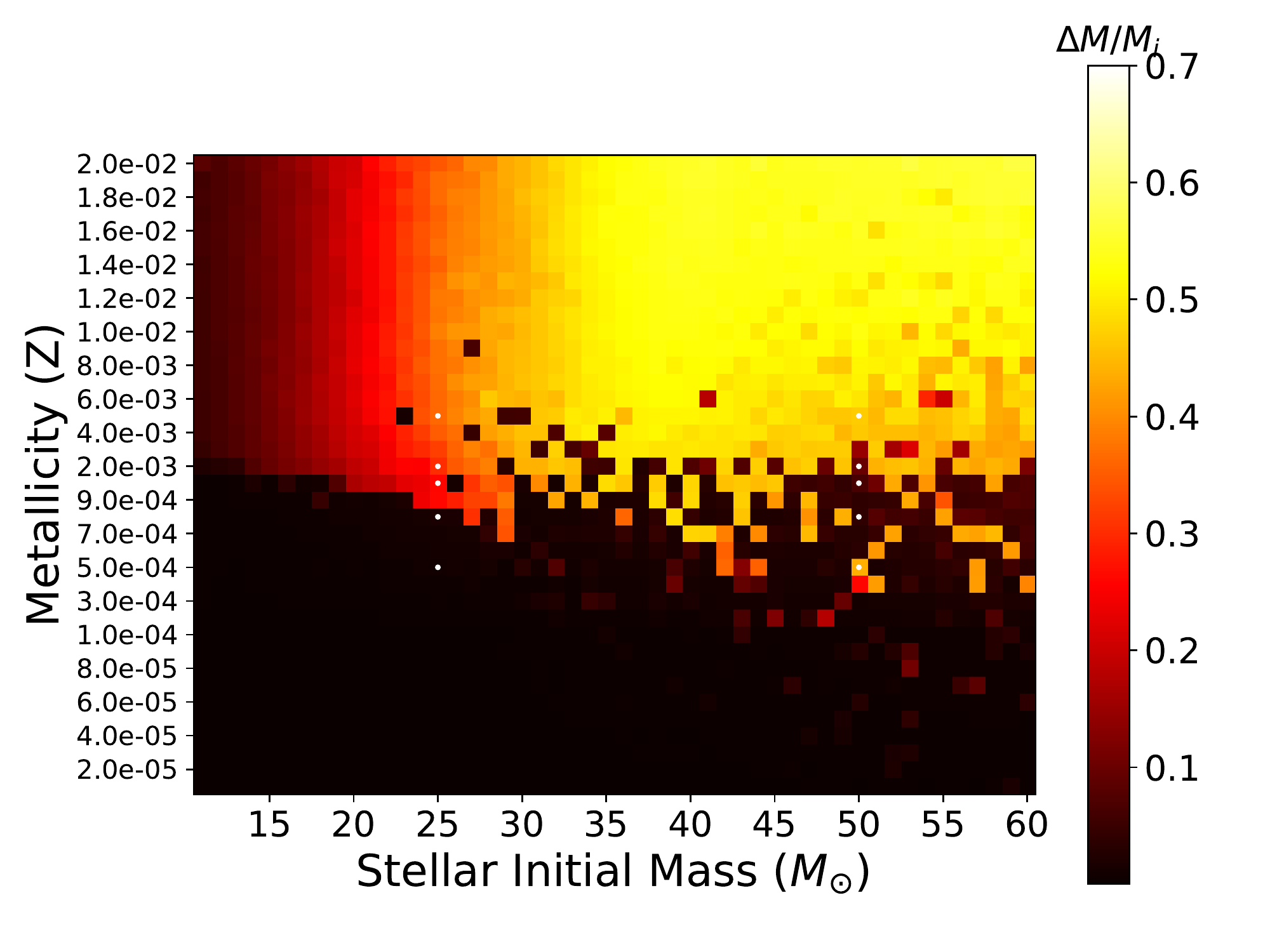}
\caption{Fraction of total mass loss in each of our 1900 models is shown in the initial mass–metallicity phase diagram.  The color scale represents the fraction of initial mass that is lost before a star dies. Each pixel in this map represents a stellar model. Note that the scale of the y-axis is nonuniform. A significant transition of the fraction value exists around $Z \sim  10^{-3}$, which separates the bright (high-$Z$) and dark (low-$Z$) sides of the map. This transition is called the \textit{critical metallicity}. The white dots mark the models chosen in Section 4.4 to investigate the origin of the critical metallicity.}
\label{fig:deltaM_map}
\end{figure*}

\textit{Herein, we call the $Z\sim 10^{-3}$ as the critical metallicity ($Z_{\rm c}$) for the mass loss of massive stars.} Generally, significant mass loss occurs only when $Z>Z_{\rm c}$. Nonetheless, $Z_{\rm c}$ is not a sharp boundary as shown in the stellar initial mass versus metallicity ($M_{\rm i} - Z$) diagram of mass loss fraction ($\Delta M/M_{\rm i}$) in Figure~\ref{fig:deltaM_map}, where all the 1900 models are included.  Two distinct regions are present in this diagram: the high-$Z$ region with mass loss fraction $\gtrsim$40\% and the low-$Z$ region with mass loss fraction $\lesssim$10\%. The separation of the high-$Z$ and low-$Z$ regions is clear, and the boundary at $Z\sim 10^{-3}$ is what we call $Z_{\rm c}$. For initial masses $\lesssim 30\,M_{\odot}$, the boundary between these two regions is sharp; for initial masses $\gtrsim 30\,M_{\odot}$, the boundary is fuzzy. 

\subsection{Mass Loss from Different Wind Schemes}

\begin{figure}[tbh]
\centering
\includegraphics[scale=0.4]{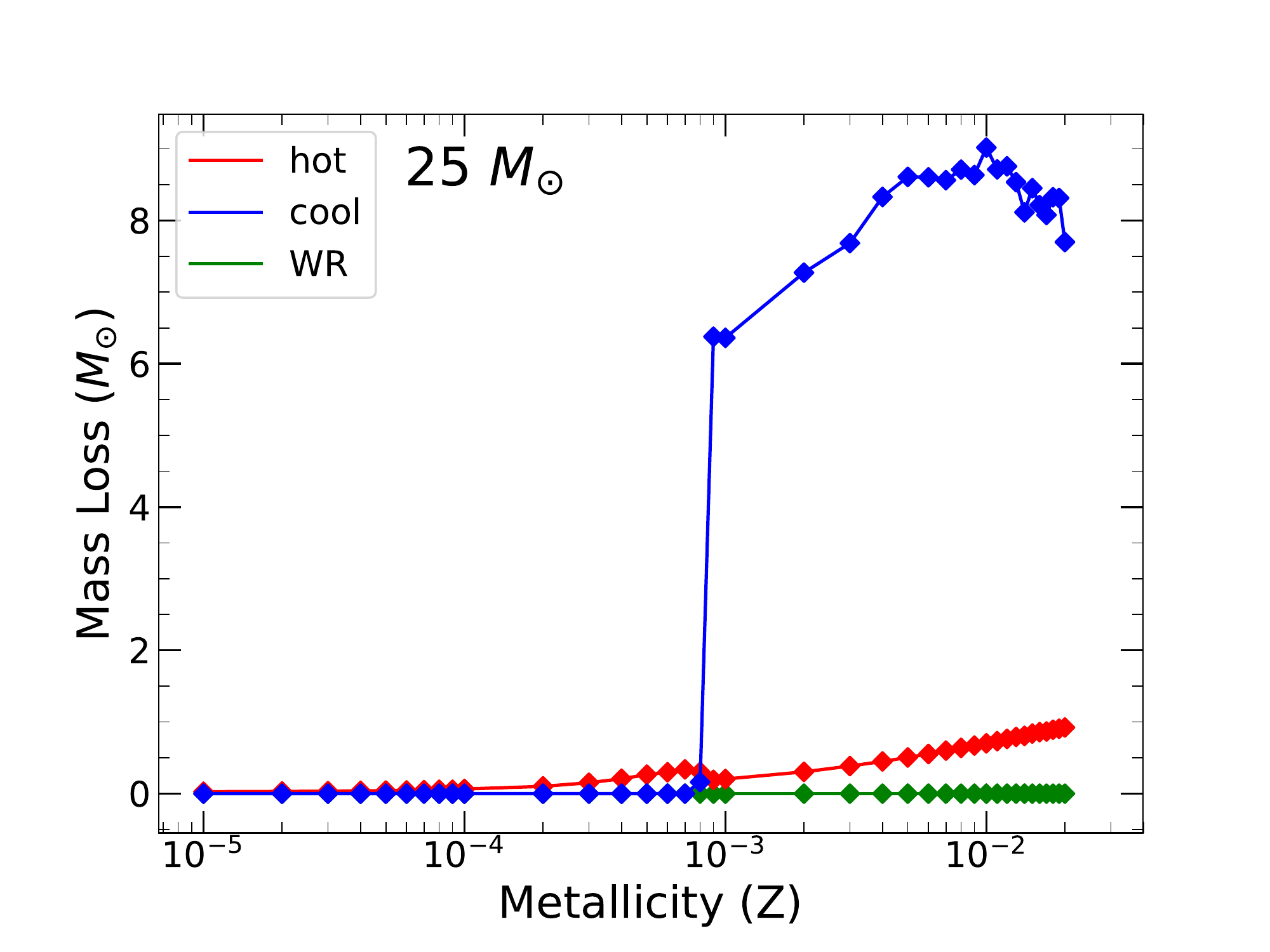}
\includegraphics[scale=0.4]{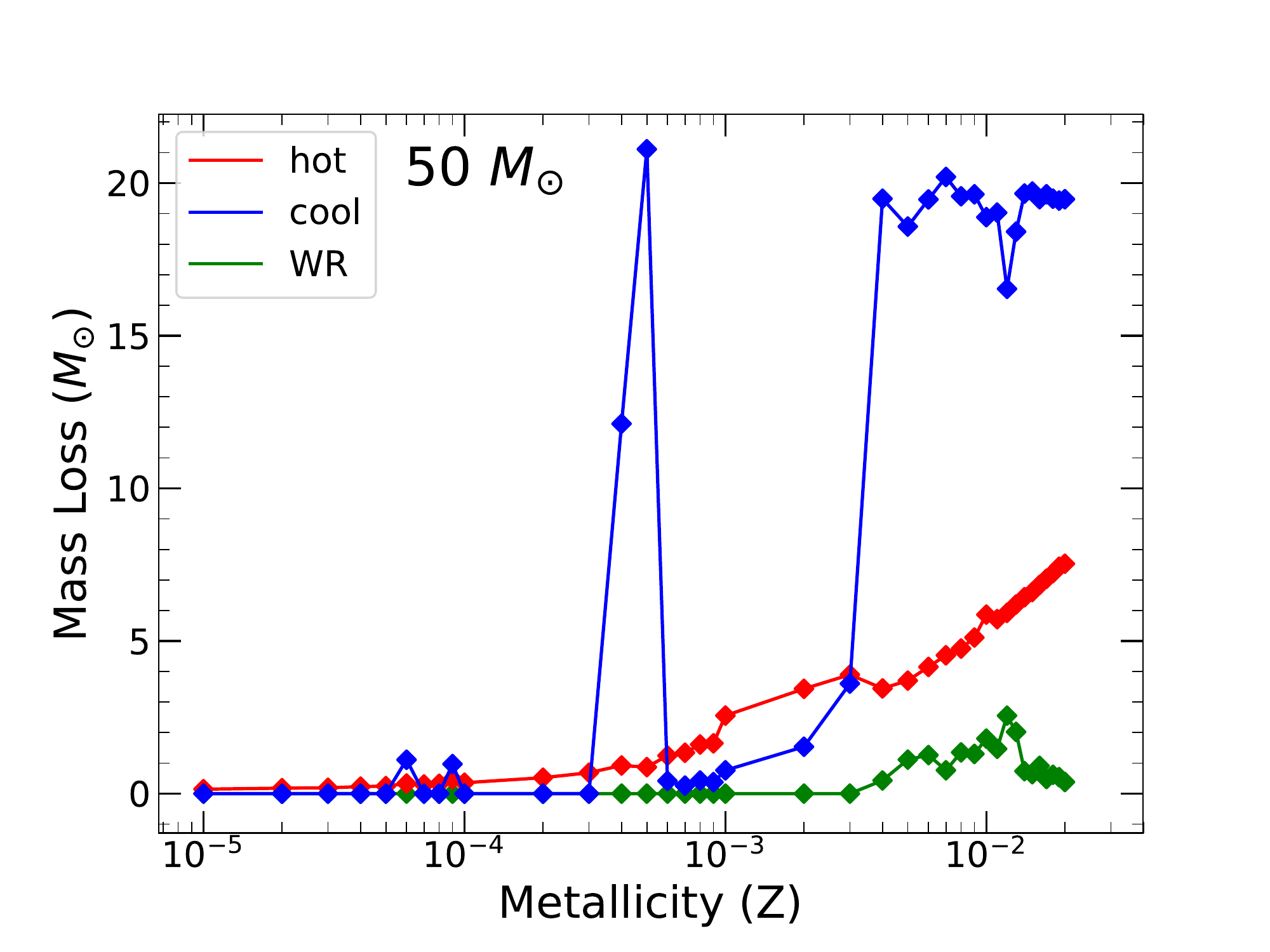}
\caption{Contributions of the hot, cool, and WR winds to the total mass loss of 25 and 50 $M_{\odot}$ stars. The significant jump in mass loss at $Z\sim 10^{-3}$ mainly arises from cool winds.}
\label{fig:hot-cool-WR}
\end{figure}

To understand the origin of $Z_{\rm c}$, we compare the individual contributions of mass loss by hot, cool, and WR winds in order to identify a specific stellar evolutionary stage that is most accountable for $Z_{\rm c}$. Figure~\ref{fig:hot-cool-WR} displays the contributions of different winds for initial stellar masses of 25 and $50\,M_{\odot}$. In both cases, a distinct jump in the mass loss occurs only in the cool wind near $Z_{\rm c}$, indicating that the cool wind is responsible for $Z_{\rm c}$.

\begin{figure}[tbh]
\centering
\includegraphics[scale=0.4]{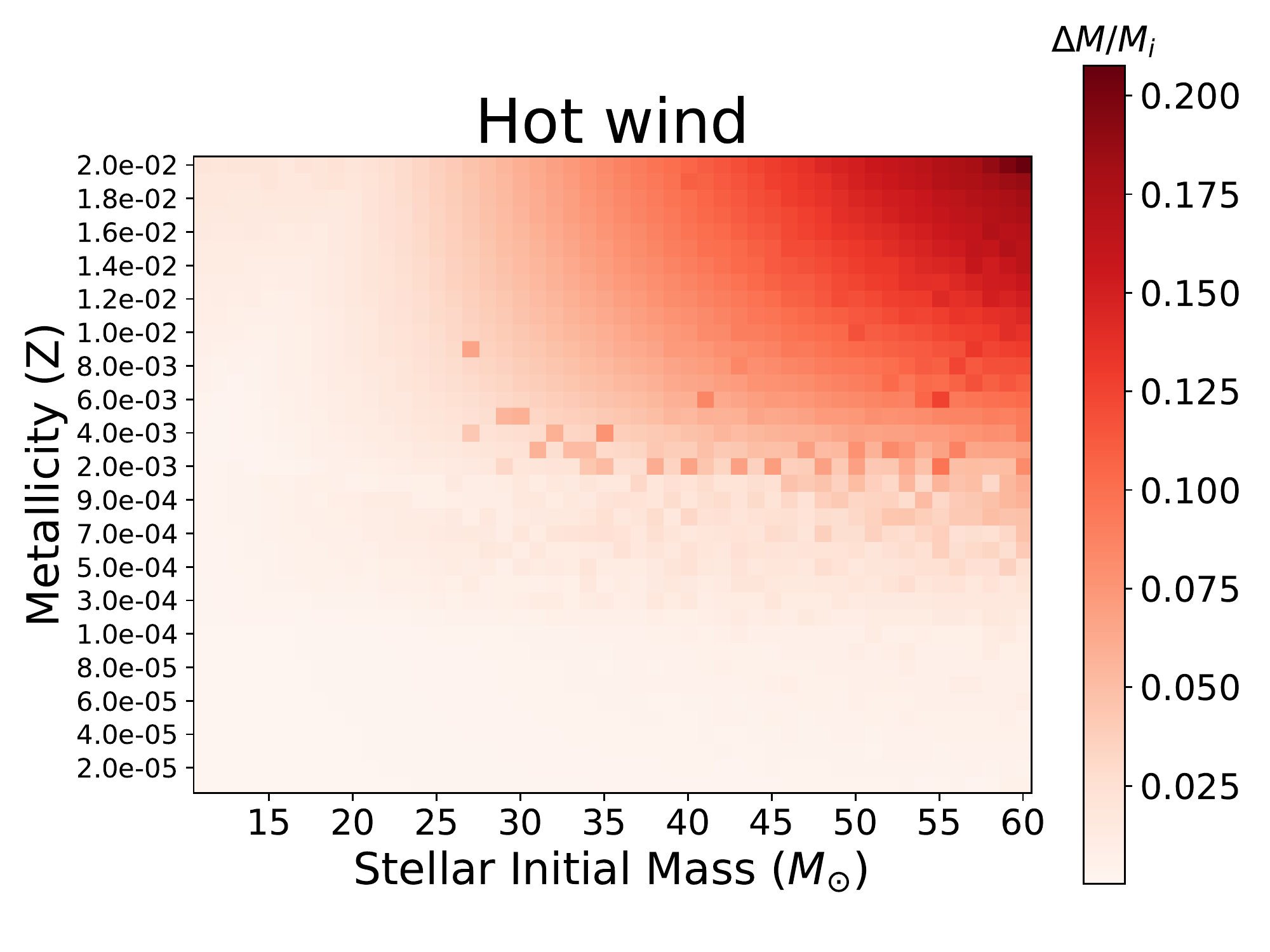}
\includegraphics[scale=0.4]{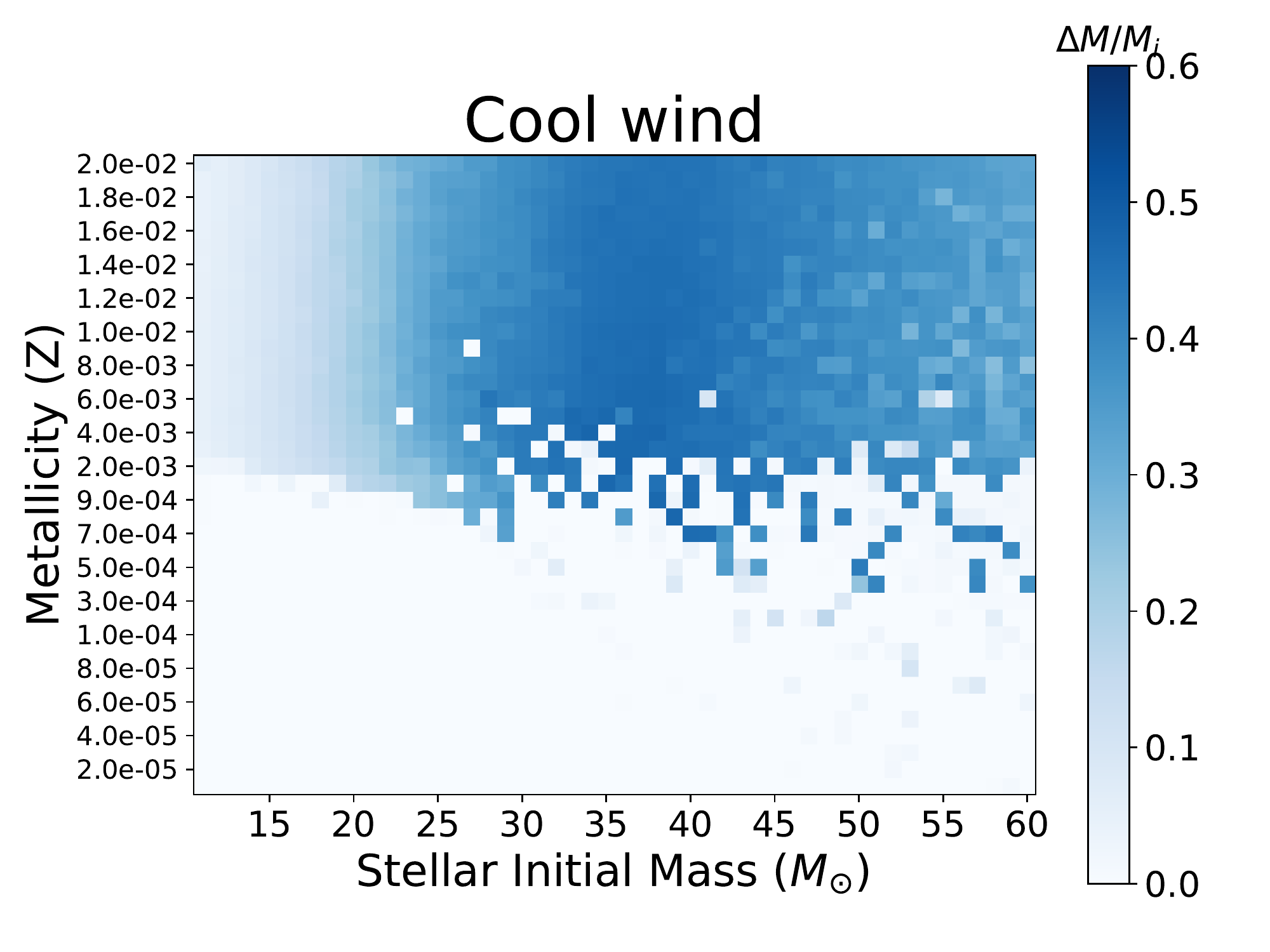}
\includegraphics[scale=0.4]{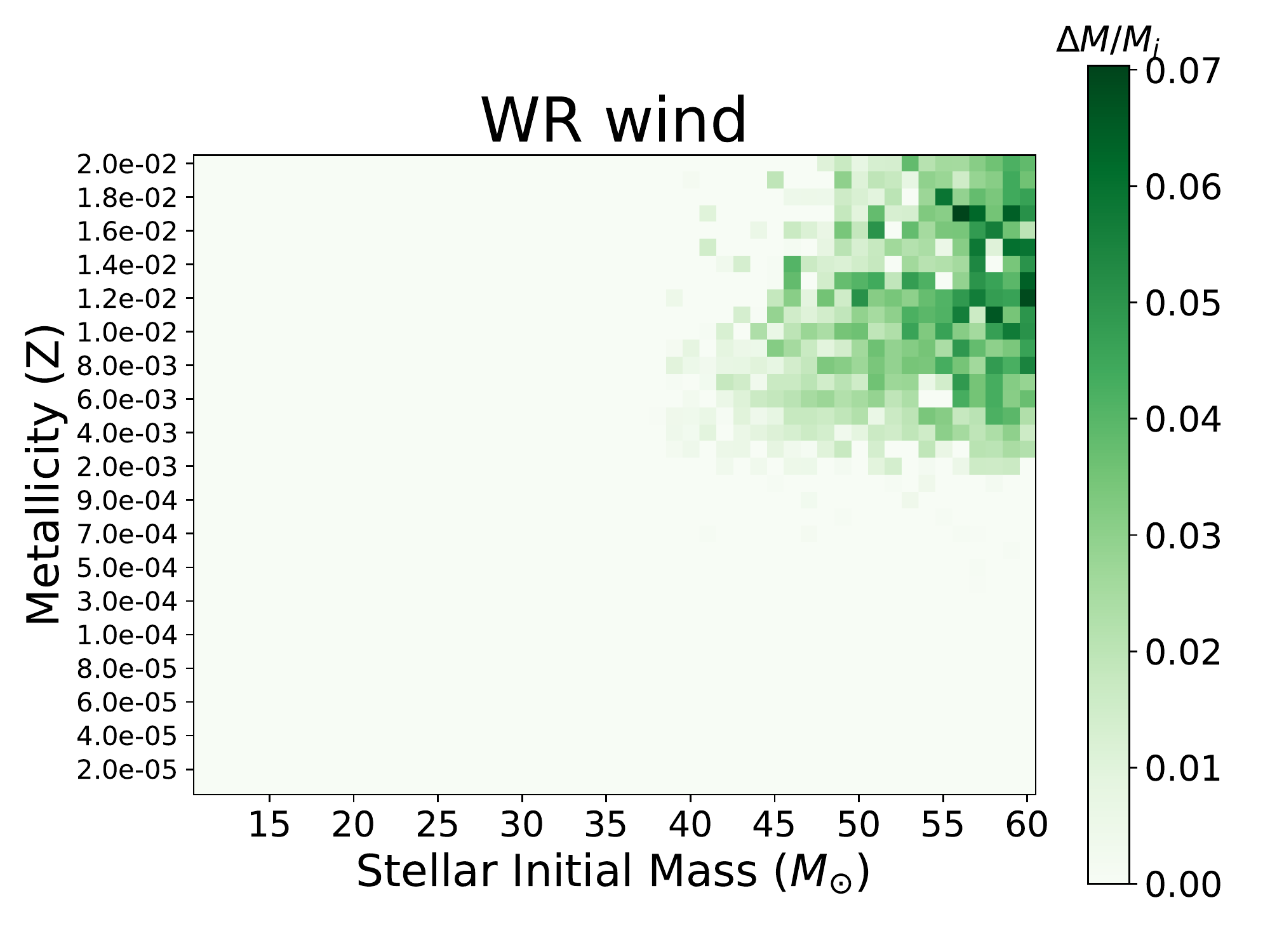}
\caption{Fractions of mass loss to the initial masses caused by the hot, cool, and WR winds. These diagrams are similar to Figure~\ref{fig:deltaM_map}, but the three wind components are separately plotted.}
\label{fig:map_hot-cool-WR}
\end{figure}

The $M_{\rm i}-Z$ diagrams of mass loss fraction for hot, cool, and WR winds are presented in Figure~\ref{fig:map_hot-cool-WR}. For cool winds, going from high $Z$ to low $Z$, the high mass loss makes a sharp transition into a low mass loss near a threshold $Z$ of $\sim 1\times 10^{-3}$.  The threshold $Z$ is slightly lower for stars with $M_{\rm i}=25-30\,M_\odot$ and higher for the more massive stars. For hot winds, the transition of high mass loss to low mass also occurs near a threshold $Z$ of $1\times 10^{-3}$, but the transition is gradual. For WR winds, the threshold $Z$ is higher, near $3\times 10^{-3}$.  As the total contribution of hot and WR winds to the mass loss is small, $\Delta M/M_{\rm i} \lesssim 20\%$, the $M_{\rm i}-Z$ diagram of cool wind is very similar to that of the total mass loss (all winds combined) in Figure~\ref{fig:deltaM_map}; furthermore, the $Z_{\rm c}$ of the total mass loss is similar to the threshold $Z$ of the cool winds.  It should be noted that at low $Z$, the hot winds may lose more mass than the cool winds up to the threshold $Z$ of the cool winds; however, the mass lost via hot winds is so much smaller than the mass lost via cool winds that the cool winds' threshold $Z$ dominates.  

Therefore, $Z_{\rm c}$ of mass loss mainly stems from cool winds. This result is reasonable because the mass-loss rates of cool winds are usually considerably higher than those of hot winds, as included in the wind prescriptions (Section \ref{sec:mdot_evolution}). 

\begin{figure*}[tbh]
\plottwo{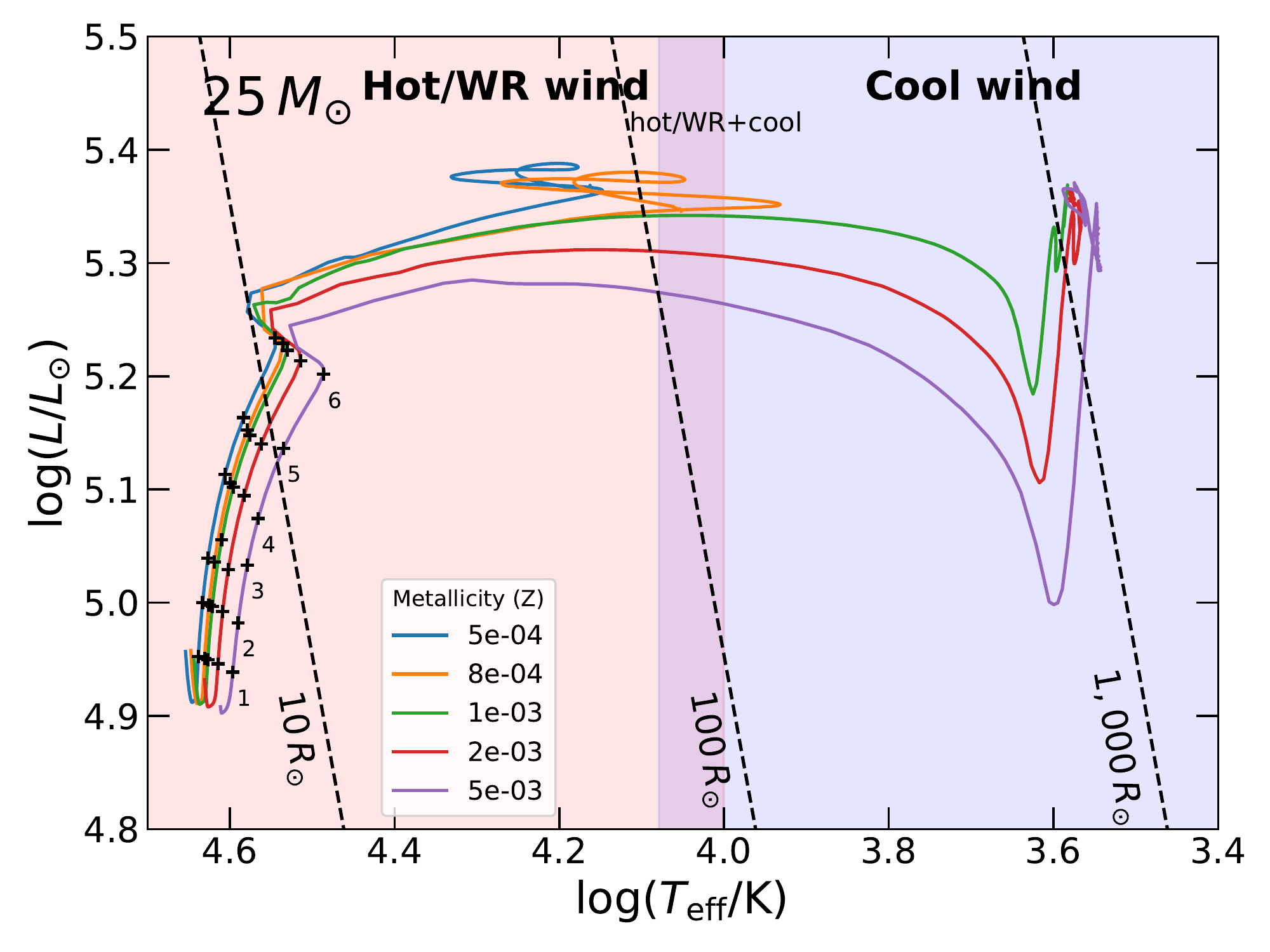}{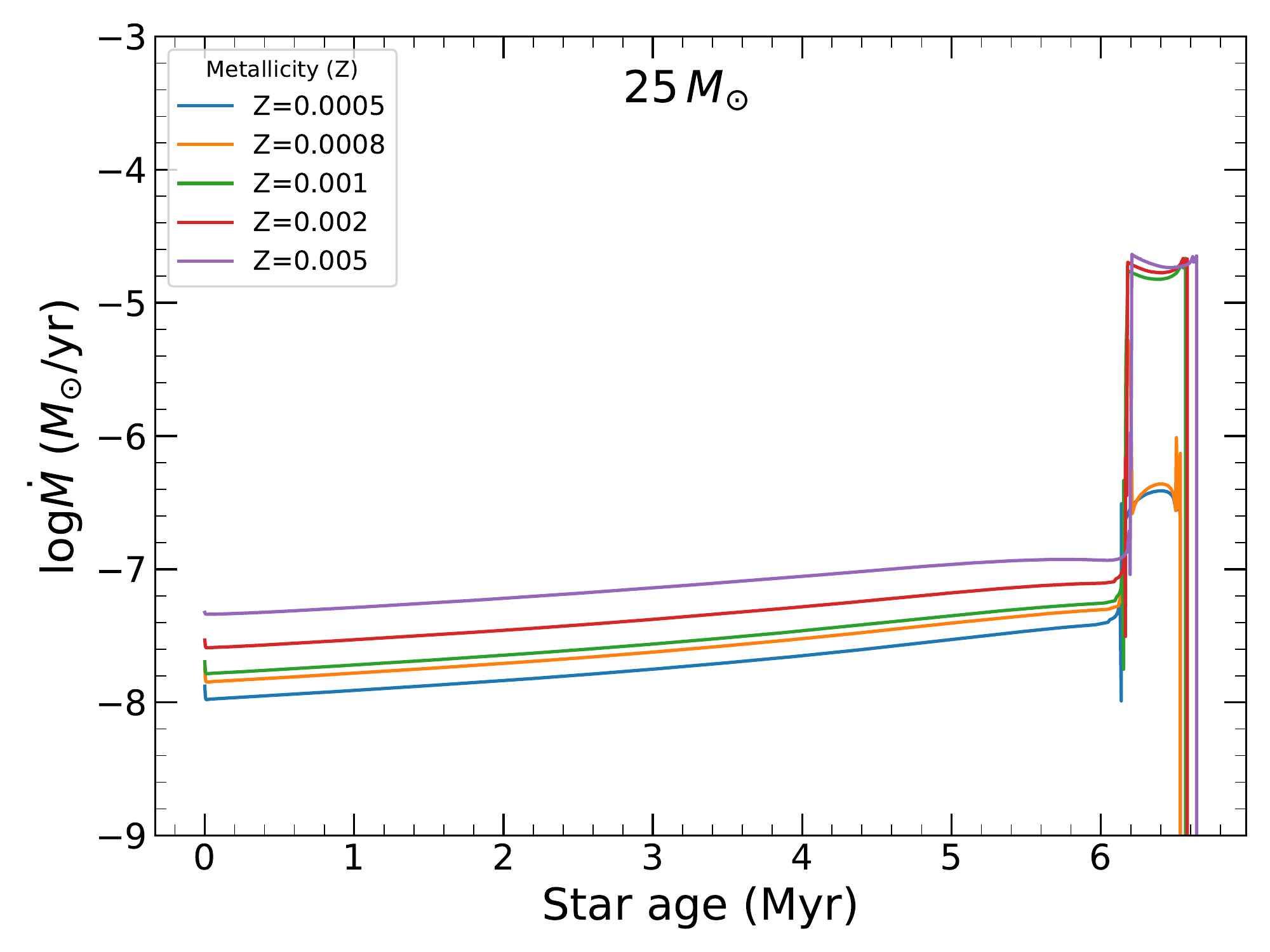}
\plottwo{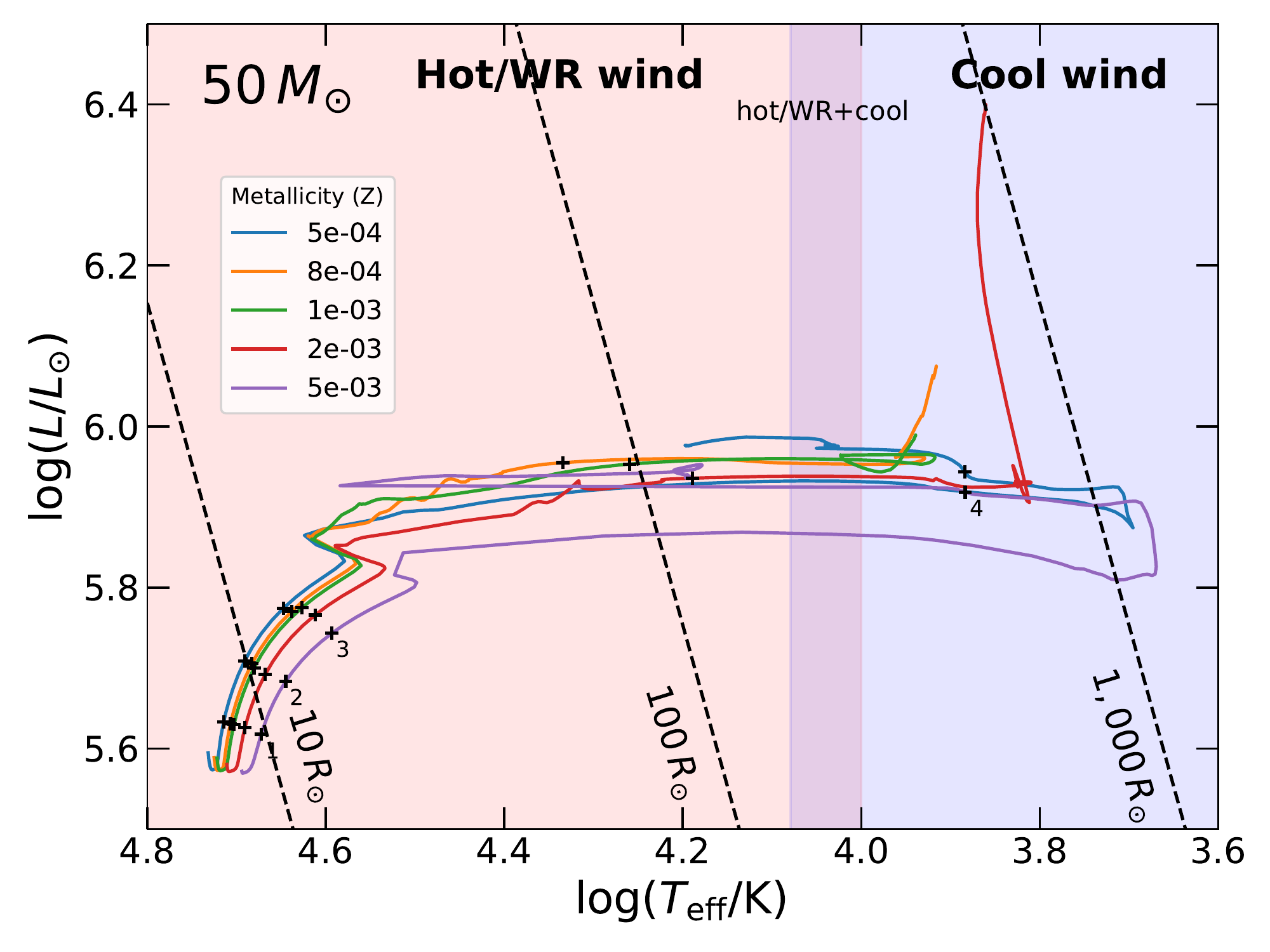}{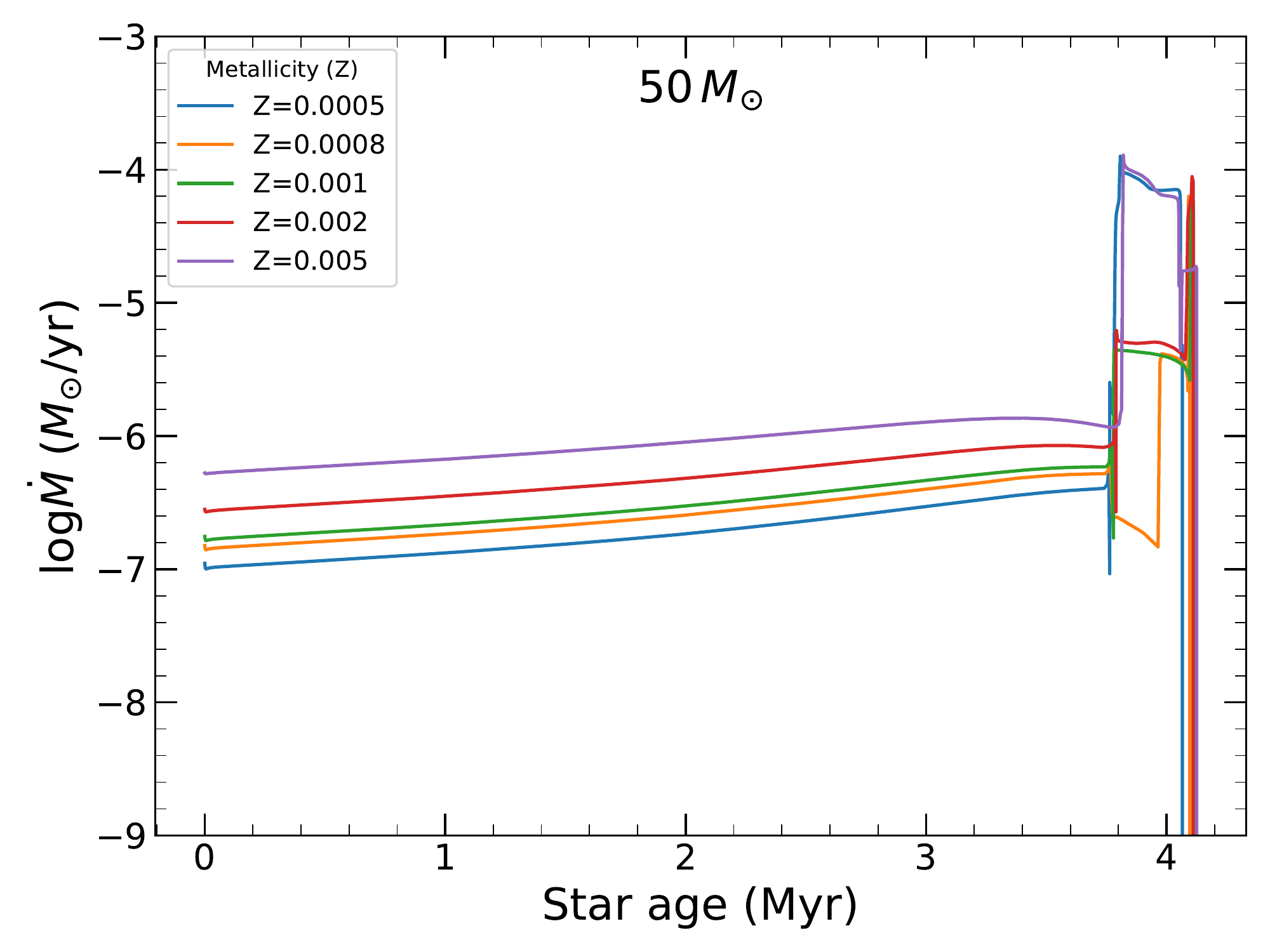}
\caption{Evolutionary tracks (\textit{left panel}) and mass-loss rates  (\textit{right panel}) of 25 and 50 $M_{\odot}$ stars with different $Z$ that are all near the critical metallicity. The bimodal behavior of the mass loss corresponds to two different trends of evolutionary tracks; the high-mass-loss models are those that successfully evolve into cool supergiants at the beginning of the post-main-sequence stage.}
\label{fig:HR-Z}
\end{figure*}

\begin{figure*}[tbh]
\includegraphics[width=\columnwidth]{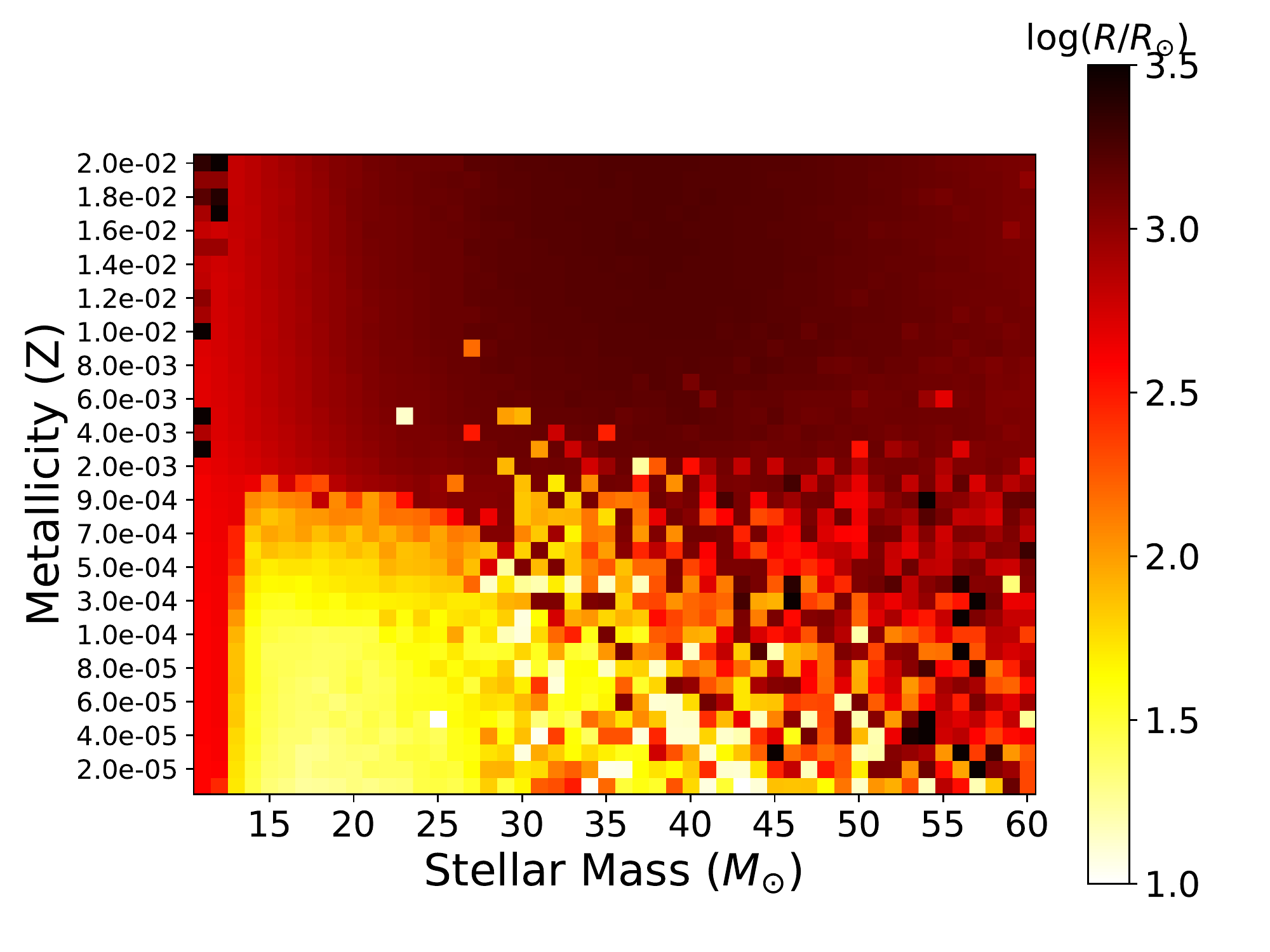}
\includegraphics[width=\columnwidth]{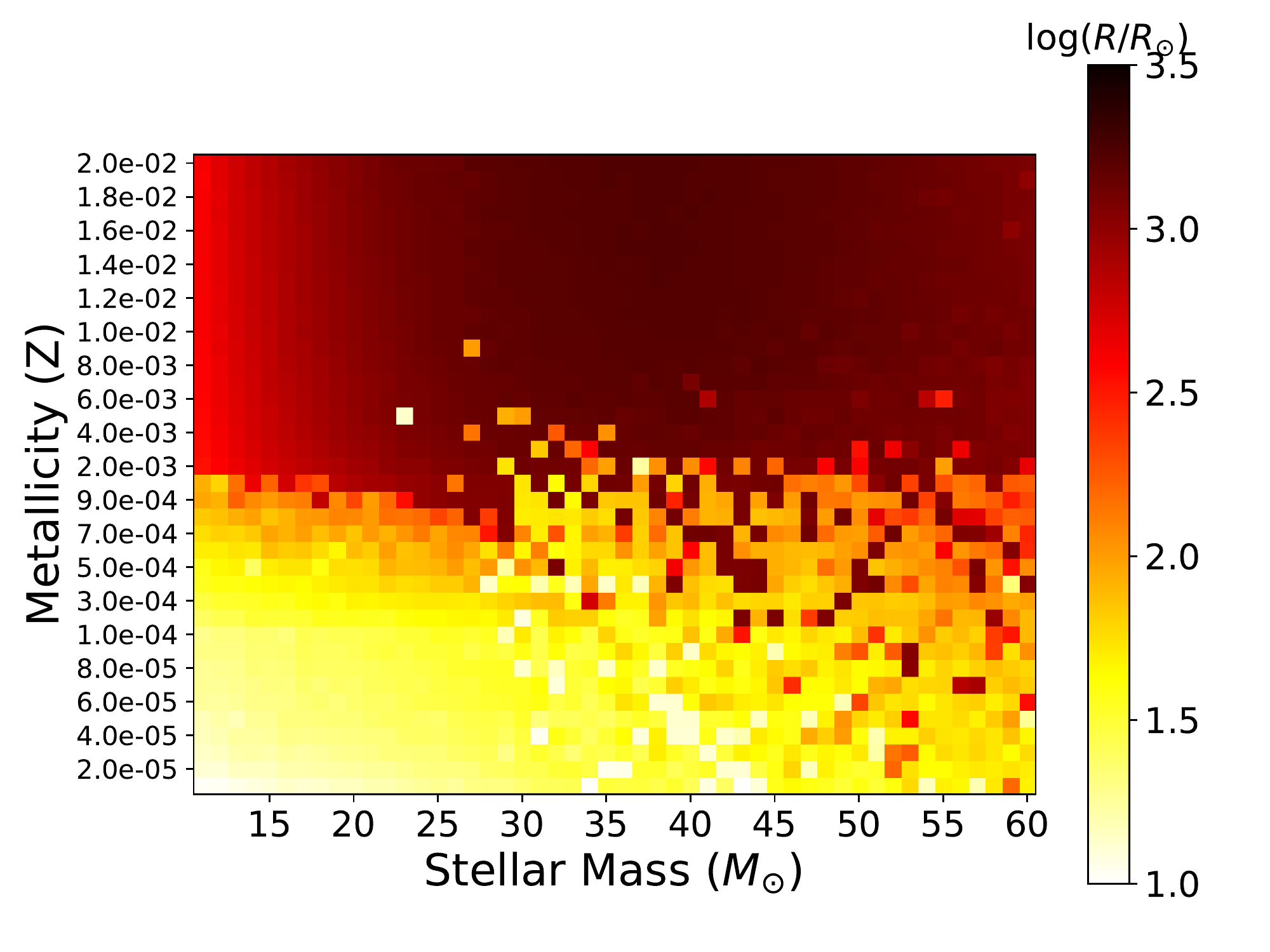}
\caption{Maximum stellar radius of each model in the entire stellar lifetime (\textit{left panel}) and before the core-carbon burning commences (\textit{right panel}). The distribution of values in the right panel is almost identical to that in Figure~\ref{fig:deltaM_map}, indicating that the low-$Z$ models with negligible mass loss are those that fail to expand to cool supergiants before the core-carbon burning commences.}
\label{fig:R-map}
\end{figure*}

\subsection{Cool Supergiants and the Critical Metallicity}

We have shown that the mass-loss fraction ($\Delta M /M_{\rm i}$) of a massive star is high for high $Z$ and low for low $Z$, making a transition at $Z_{\rm c}$ that is caused mainly by the cool winds. To gain more insights into the dichotomy of mass-loss fraction and its scatter around $Z_{\rm c}$, we use 25 $M_{\odot}$ and 50 $M_{\odot}$ stars with different metallicities that are near $Z_{\rm c}$ as examples, and look into their post-main-sequence evolution. For each stellar mass, five models with $Z$ = 0.0005, 0.0008, 0.001, 0.002, and 0.005 are examined.  These models are marked by white dots in Figure~\ref{fig:deltaM_map}.  It can be seen that the 25 $M_\odot$ stars show monotonic increase of mass loss fractions with increasing $Z$, while the 50 $M_\odot$ stars show non-monotonic variations of mass loss fraction across $Z_{\rm c}$. 
We deliberately choose these two masses with contrasting mass loss variations across $Z_{\rm c}$, and expect the comparison between them may shed light on the physical origin of the transition and the scatter of mass loss fraction near $Z_{\rm c}$.



To investigate the crucial differences among these models, we plot their stellar evolutionary tracks and mass-loss rates in Figure~\ref{fig:HR-Z}.  For the 25 $M_{\odot}$ stars, the three models yielding high $\Delta M$ ($Z$ = 0.001, 0.002, and 0.005) are those that evolve to radii $R>1,000\, R_{\odot}$ and $\log(T_{\textrm{eff}}/\textrm{K}) < 3.6$ (i.e., $T_{\textrm{eff}} < 4,000$ K). In other words, these stars end their lives as RSGs. Due to stellar expansion, their mass-loss rates increase by 2--3 orders of magnitude in the supergiant stages. In contrast, the two models yielding in low $\Delta M$ ($Z=0.0005$ and $0.0008$) only expand to $\sim$100 $R_{\odot}$. After moderate expansions, they bounce back to the left (i.e., higher $T_{\textrm{eff}}$) in the HR diagram, and their $T_{\textrm{eff}}$ remains higher than $\sim$10,000 K. Such a high $T_{\textrm{eff}}$ prevents the cool wind from operating, and thus, their mass-loss rates only slightly increase during the post-main-sequence stages. In summary, the two different behaviors of stellar expansion result in the dichotomy of the total mass loss.

For the 50 $M_{\odot}$ stars, the evolutionary tracks are more complicated. At the age of 4 Myr (marked by “+” in the HR diagram in Figure~\ref{fig:HR-Z}), the two models yielding high $\Delta M$ ($Z=0.0005$ and $0.005$) have already evolved into the cool-wind domain, and eventually expand to $R>1,000\, R_{\odot}$. On the other hand, the three models yielding low $\Delta M$ ($Z=0.0008$, $0.001$, and $0.002$) are still in the hot-wind domain at 4 Myr. These three models eventually enter the cool-wind domain at the very end of their lifetimes when the core-carbon burning commences, and approach a high mass-loss rate of $\sim 10^{-4}$ $M_{\odot}$ yr$^{-1}$. However, their cool winds only last a few $\times 10^4$ yr. Thus, the integrated $\Delta M$ is still small, yielding $\Delta M/M_i < 10\%$.

The above examples demonstrate that the crucial process that determines the total mass loss is the expansion toward the formation of a supergiant. If a star expands to a cool supergiant with $R>1,000\, R_{\odot}$ at the beginning of the post-main-sequence stage, significant mass loss will occur. Otherwise, if a star never expands to the extent of a cool supergiant or only expands to that extent at a very 
late stage (i.e., after the core-carbon burning begins), only a minimal amount of mass loss will occur.

The crucial role played by the supergiant expansion in the dichotomy of mass loss is further illustrated in Figure~\ref{fig:R-map}, which presents maps of maximum radii in the $M_{\rm i}-Z$ grid.  We have considered maximum radii over two time spans: the entire lifetime (left panel of Figure~\ref{fig:R-map}) and the time span from the ZAMS to the beginning of core carbon burning (right panel of Figure~\ref{fig:R-map}). 
The left panel shows a complex map with a visible dichotomy only for stellar masses of 14--27\,$M_{\odot}$. The complexity stems from the maximum radii arising from different evolutionary stages and paths for stars of different masses. For high metallicities ($Z>Z_{\rm c}$), stars reach maximum radii during the core helium burning; while for low metallicities ($Z<Z_{\rm c}$), 14--27\,$M_{\odot}$ stars reach the maximum radii during the core helium burning, but the higher mass stars reach maximum radii after the core carbon burning.
The core carbon burning stage is pretty late in a massive star's lifetime, and the duration for high mass loss rate is short; thus, massive stars reach maximum radii during the core carbon burning stage have small mass loss fractions.  The complications at the core carbon burning stage makes the maximum radius over the entire lifetimes an ineffective indicator of total mass loss.  

As the core helium burning stage lasts the longest after the main sequence, and the bulk of mass loss occur during this stage, we next consider the maximum stellar radii during the time span from the ZAMS to the beginning of core carbon burning.  In the right panel of Figure~\ref{fig:R-map}, a cleaner dichotomy is seen and the transition occurs near $Z_{\rm c} \sim 0.001$.  In fact, this map of maximum stellar radii resembles closely the map of mass loss fraction (Figure~\ref{fig:deltaM_map}): both show dichotomy across the $Z_{\rm c} \sim 0.001$ and larger scatter around $Z_{\rm c}$ at high stellar masses.  Examined closely, even the scatter exhibits one-to-one correspondences between these two maps. 
These similarities indicate that the maximum stellar radius before core carbon burning is an effective indicator of total mass loss.

We further find that models with high mass loss are almost exactly those that have expanded to cool supergiants with $R>1,000\,R_{\odot}$ at the core-helium burning stage, and these models are usually of $Z\gtrsim 10^{-3}$. 
Therefore, we conclude that the $Z_{\rm{c}}$ of mass loss is essentially the \textit{critical metallicity of cool supergiant formation}. 

Generally, cool supergiant formation and the corresponding mass loss only occur when $Z$ is higher than the critical metallicity ($Z_{\rm{c}} \sim 10^{-3}$). 
For stars with $Z>Z_{\rm{c}}$, they usually expand to $R>1,000\,R_{\odot}$ and become cool supergiants at the beginning of the post-main-sequence stage. These stars can lose large amounts of mass due to cool winds. 
In contrast, stars with $Z<Z_{\rm{c}}$ either do not reach $R>1,000\,R_{\odot}$ throughout their lifetimes or only expand to this extent at the very final stage of stellar evolution. For these low-$Z$ stars, the cool wind scheme is not activated in most of their stellar lifetime; thus, their total mass loss is very low. 
For $Z\sim Z_{\rm{c}}$, a stars either successes or fails to become a cool supergiant, and thus show erratic variations of the total mass loss as a function of metallicity (Figure~\ref{fig:mtot_m} and \ref{fig:deltaM_map}). The physical mechanism that determines whether a star can become a cool supergiant will be investigated in our Paper II.

\section{Feedback from Stellar Winds}\label{sec:feedback}

Our grid of stellar evolution models can be used to determine the feedback from stellar winds. We first compute the kinetic energy released by winds from individual stars, then consider the wind energy feedback for a star cluster.  For star clusters we investigate the metallicity dependence of the total mass and kinetic energy injected by stellar winds into the ISM, and the evolution of mass and energy injection rates.

\subsection{Kinetic Energy Released by Winds}
The flux of kinetic energy carried by a stellar wind is called mechanical luminosity of the wind, defined as $L_{\rm w} \equiv \frac{1}{2} \dot{M}v_{\infty}^2$, where $v_{\infty}$ is the wind terminal velocity.  The determination of $\dot{M}$ has already been described in Section~\ref{sec:method}. 
Our stellar evolution models do not provide $v_{\infty}$ directly, but provide escape velocity ($v_{\rm{esc}}$) that can be scaled to determine $v_{\infty}$.  
For hot winds, we follow V01 and directly use the $v_{\infty}/v_{\rm{esc}}$ scaling relations given in Eq. A1 and A2 in Appendix \ref{sec:appA}.  
For cool winds, we follow \citet{leitherer1992} and set the RSG wind terminal velocity at a constant 30 km s$^{-1}$. More details of $v_{\infty}/v_{\rm{esc}}$ can be found in Appendix \ref{sec:appA}.
A star's $\dot{M}$ and $v_{\infty}$ are used to compute $L_{\rm w}$, then $L_{\rm w}$ is integrated over the star's lifetime to determine the total kinetic energy injected by its winds into the ISM.

\begin{figure}[tbh]
\includegraphics[width=\columnwidth]{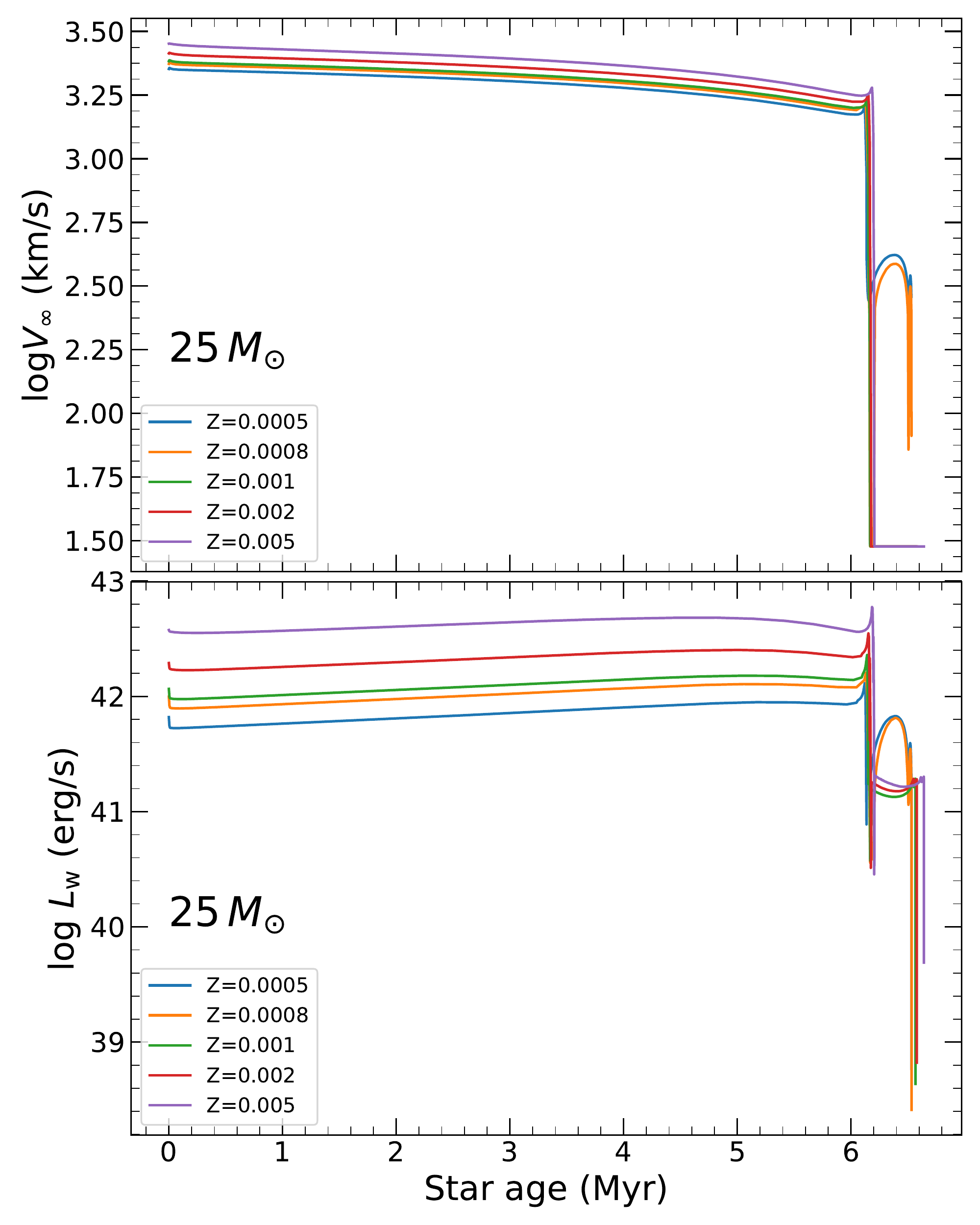}
\caption{The evolution of wind terminal velocity $v_{\infty}$ (\textit{top panel}) and mechanical luminosity $L_{\rm w}$ (\textit{bottom panel}) of stars with initial mass 25 $M_{\odot}$ and different metallicities near $Z_{\rm{c}}$. Sudden transitions of $v_{\infty}$ and $L_{\rm w}$ occur when the stars enter the post-main-sequence stage after the age of 6 Myr.}
\label{fig:KE_age}
\end{figure}

Figure~\ref{fig:KE_age} illustrates $v_{\infty}$ and $L_{\rm w}$ of 25 $M_{\odot}$ stars with different $Z$ values around $Z_{\rm{c}}$. Main-sequence O stars produce fast winds of $\sim$1500--3000 km s$^{-1}$.
When a star evolves into an RSG, its $v_{\infty}$ suddenly drops by a factor of 100. At the same time, its $\dot{M}$ goes up by 2--3 orders of magnitude (Figure~\ref{fig:HR-Z}). As a result, $L_{\rm w}$ decreases by 1--2 orders of magnitude when the star evolves from a main-sequence O star to an RSG. Although cool RSG winds release large amounts of mass, they only carry relatively small amounts of kinetic energy because of their slow wind velocities. 
For stars with higher initial masses and metallicities, the domination of hot winds in the energy output is even more significant.

\begin{figure}[tbh]
\includegraphics[width=\columnwidth]{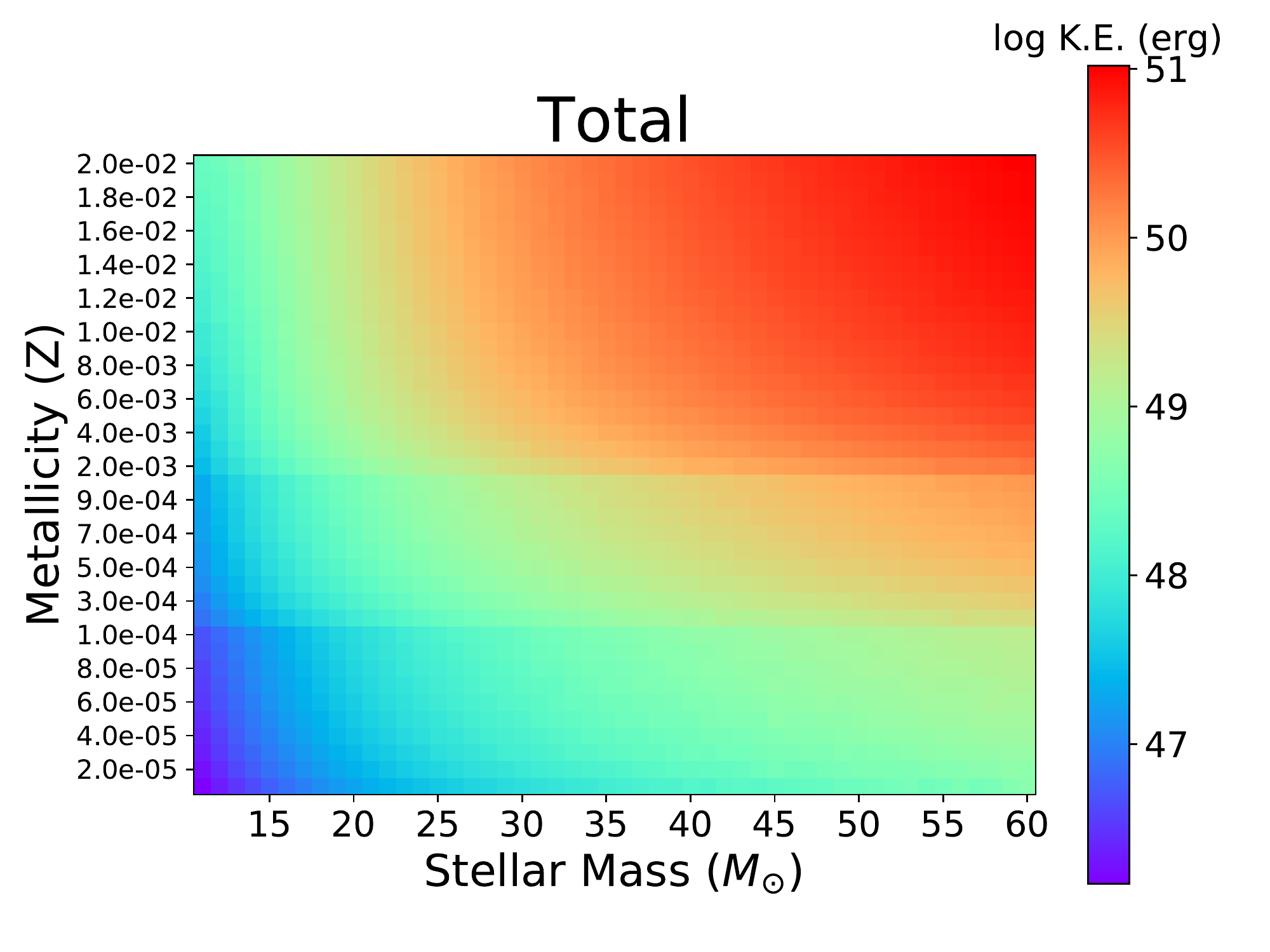}
\includegraphics[width=\columnwidth]{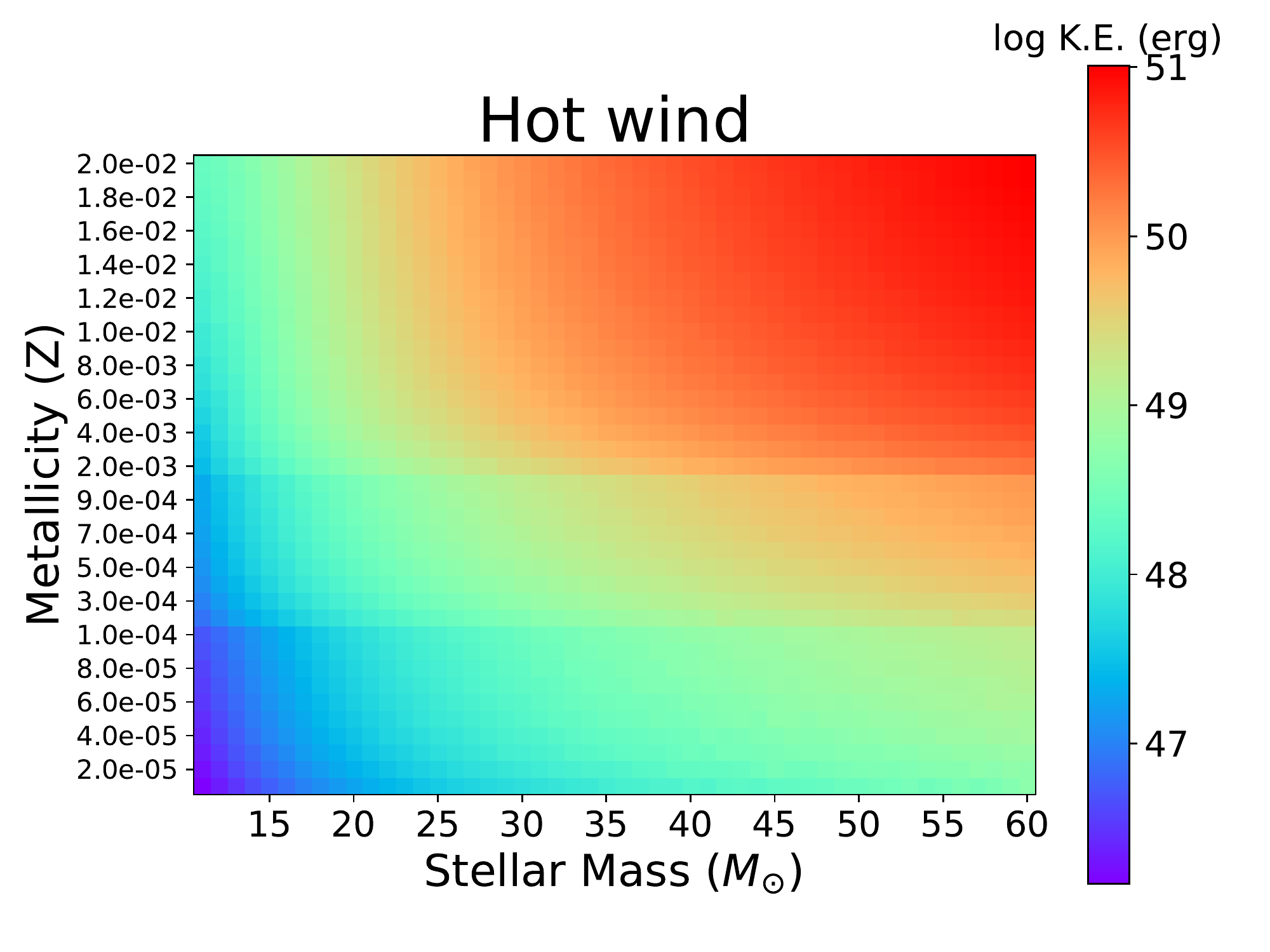}
\includegraphics[width=\columnwidth]{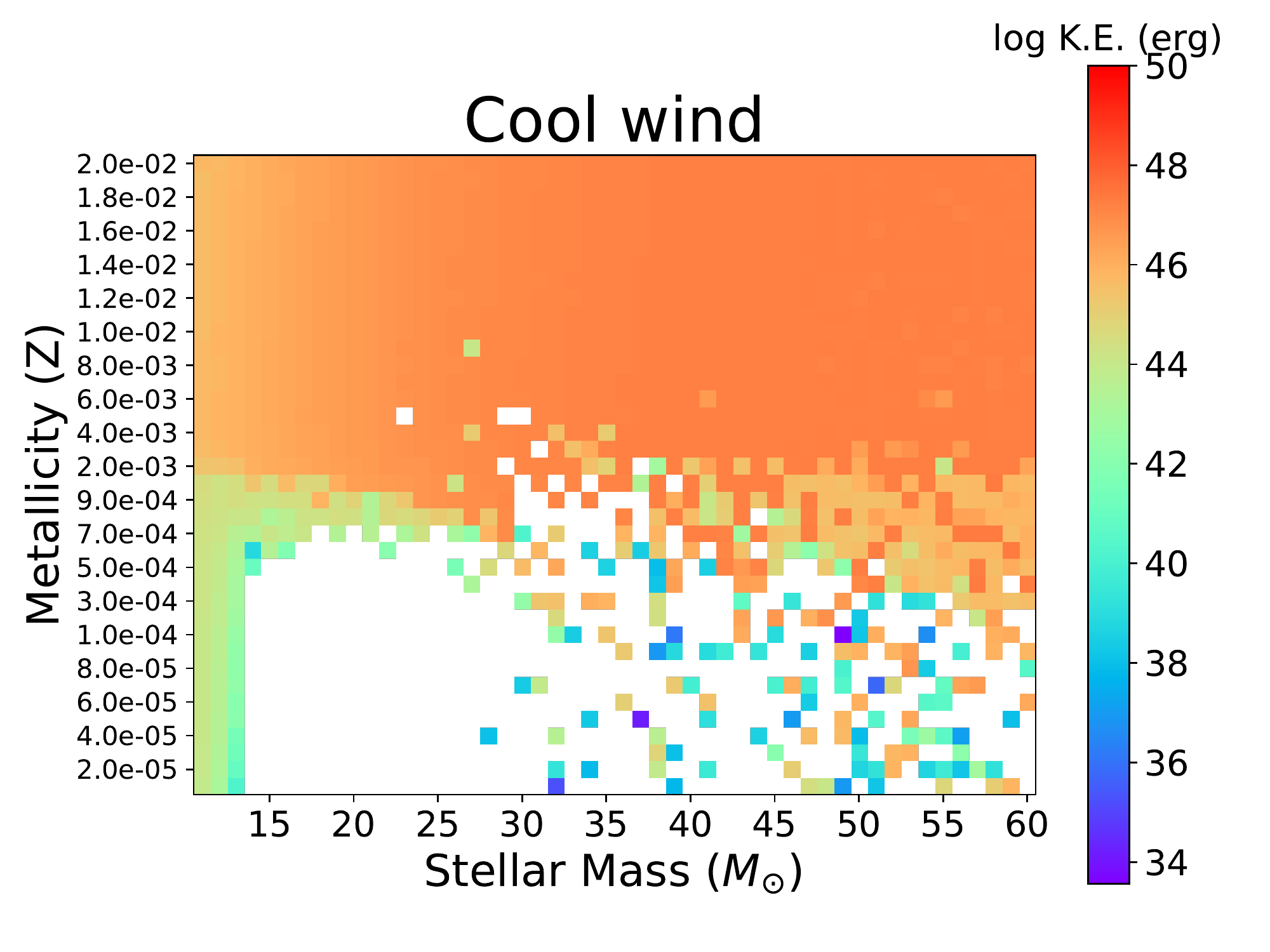}
\caption{The total kinetic energy released by stellar winds throughout the stellar lifetimes. Each pixel in this map represents a stellar model. Note that the y-axis scale is non-uniform. In addition to the total energy, we separately plot the contributions from hot and cool winds, and hot winds dominate the kinetic energy released by stellar winds.}
\label{fig:KE_map}
\end{figure}

Figure~\ref{fig:KE_map} presents the total kinetic energy output over the stellar lifetime of each model. This figure also shows the contributions of the hot and cool winds separately. It is evident that the hot winds dominate the energy output, while the contributions from cool winds are almost negligible. The dominance of hot winds is attributed to their velocities being much faster than the cool winds' velocities, and their duration (including the main-sequence stage) spanning $\sim$ 90\% of the stellar lifetime. It can also been seen from the top panel of Figure~\ref{fig:KE_map} that the most massive stars with near-solar metallicities each inject $10^{50}-10^{51}$ erg of kinetic energy into the surrounding ISM, which is comparable to the explosion energy of a SN. 
In contrast, the cool winds of such a massive star contribute only up to $\sim 10^{47}$ erg to the total kinetic energy.

It is interesting to note that the $Z_{\rm c} $ for the dichotomy of total mass loss show some effects only the cool winds' total kinetic energy.  This is understandable because $Z_{\rm c}$ is closely associated with the onset of cool winds, as discussed in Section 4.3.  The stellar winds' total kinetic energy injected into the ISM is dominated by contributions from hot winds.  As cool winds play a negligible role in the total wind energy, its closely associated $Z_{\rm c}$ does not affect the total wind energy or total hot wind energy.
In other words, whether a star becomes an RSG has no bearing on the total wind kinetic energy injected to the ISM.

\subsection{Wind Feedback from a Star Cluster}
The mass loss and kinetic energy output from stellar winds of individual stars can be used to evaluate the feedback from a star cluster.  We consider a star cluster from a single burst of star formation with an initial mass function (IMF) $\xi(M) = \xi_0 M^{-\alpha}$, where $M$ is the initial stellar mass, $\xi_0$ is a scaling factor, and $\alpha$ is a constant. Here, $\xi_0$ is scaled by the number of stars in the system $N$ so that $dN = \xi_0 M^{-\alpha} dM$. In our calculations, we apply $\alpha = 2.35$ \citep[the Salpeter IMF;][]{salpeter1955}. For comparison, the widely used IMF in \citet{kroupa2001} gives $\alpha = 2.3$ for stars with masses $>1\,M_{\odot}$, which is similar to the Salpeter IMF.

We have calculated the feedback for a cluster with a total mass of $10^5$ $M_{\odot}$. Although this mass is associated with a super star cluster, the feedback scales linearly with the cluster mass and can be used to determine feedback of clusters of any other masses. For the cluster members, we adopt the canonical initial stellar mass range of $1-150\,M_{\odot}$. We note that the upper mass limit is not important because the population of very massive stars is much smaller than that of lower-mass stars. With the above assumptions, we integrate the $\dot{M}$ and $L_{\rm w}$ of individual stars to calculate the wind feedback from the entire cluster. The feedback we present only include the contributions from massive stars of $M_{\rm i}=11-60\, M_{\odot}$, which is the mass range of our simulations; thus, our results show the lower limit of wind feedback.

\begin{figure*}[tbh]
\includegraphics[width=\textwidth]{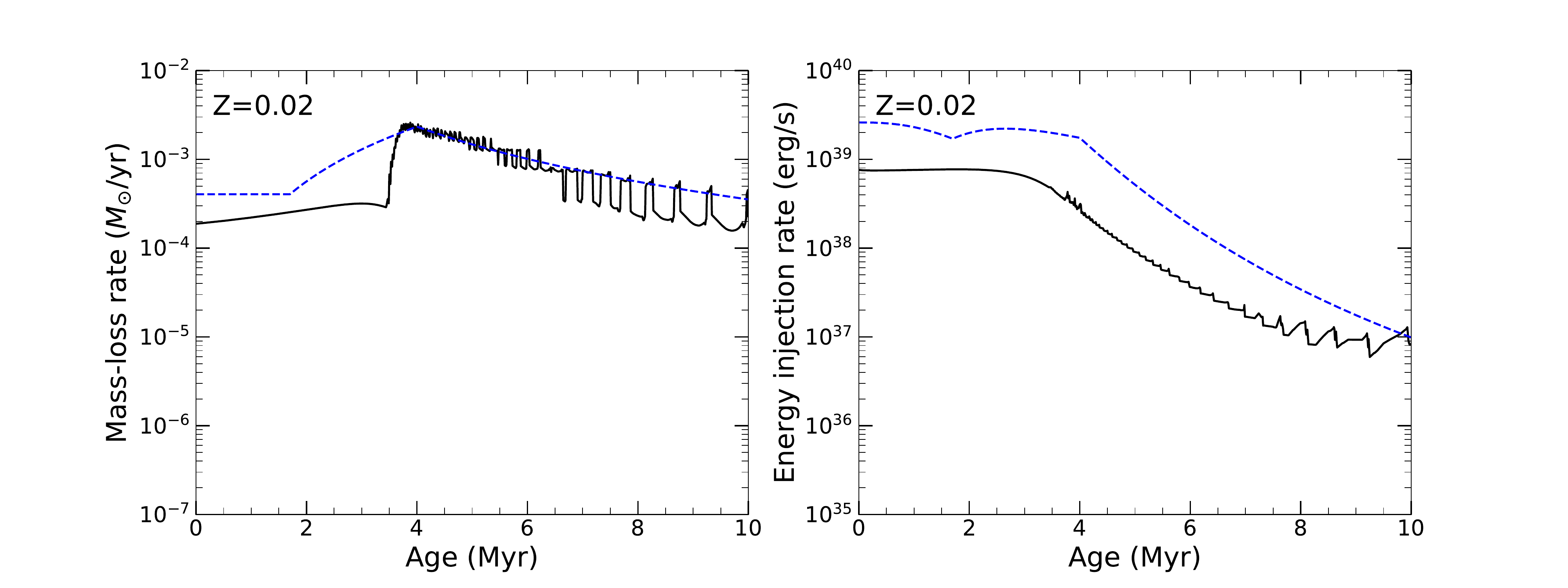}
\includegraphics[width=\textwidth]{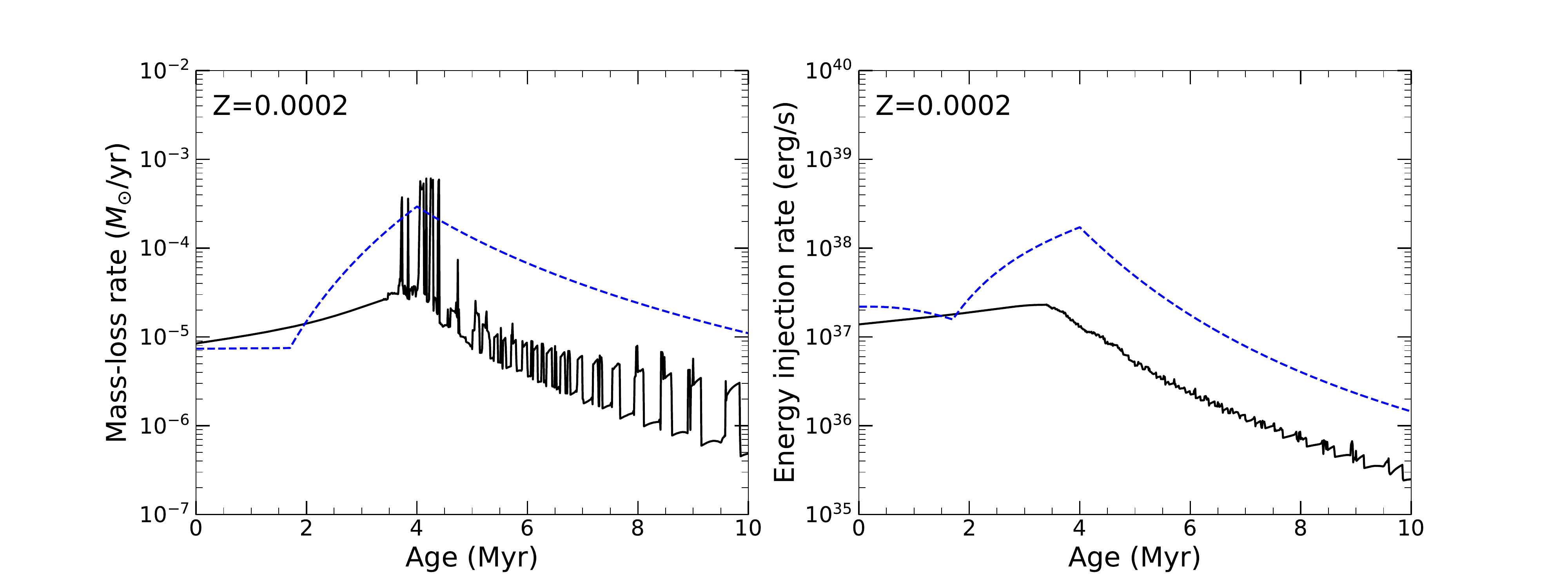}
\caption{The integrated mass-loss rate and energy-injection rate from stellar winds in a star cluster of $10^5$ $M_{\odot}$, in which the Salpeter IMF is assumed. The upper and lower panels show the results of a $Z=0.02$ cluster and a $Z=0.0002$ cluster, respectively. The blue dash lines show the functions adopted in the feedback treatment of the FIRE-3 cosmological simulation \citep{hopkins2022}.}
\label{fig:cluster_mdot}
\end{figure*}

In Figure~\ref{fig:cluster_mdot}, we choose a high-$Z$ ($Z=0.02$) case and a low-$Z$ ($Z=0.0002$) case and plot their integrated mass-loss rates and energy-injection rates of a 10$^5$ $M_{\odot}$ cluster before the age of 10 Myr. At about 3--4 Myr, the most massive stars evolve into the post-main-sequence stage. In the high-$Z$ cluster, these massive stars successively become cool supergiants and significantly enhance the overall mass-loss rate. However, the lifetimes of cool supergiants are short, so each star can only contribute to the enhancement of mass-loss rate for a short time. When stars of lower masses become cool supergiants, the more massive stars have already ended their lifetimes. The overall trend of the mass-loss rate is a gradual decrease after the sudden enhancement at about 3--4 Myr. The jagged appearance of the curve of mass-loss rate is caused by the limit of mass resolution (1 $M_{\odot}$) in our simulations.
For the low-$Z$ cluster, the stars do not evolve into cool supergiants at the helium-core burning stage, so the mass-loss rate after 4 Myr can only increase to $~10^{-5}$ $M_{\odot}$/yr due to the bi-stability jumps. Some sudden spikes of mass-loss rate occur at about 4 Myr because some low-$Z$ stars reach cool supergiants in the very late stage of their lifetimes, and thus they have enhanced the mass-loss rate for very short times. 

While the mass-loss rate of the high-$Z$ cluster is enhanced by cool winds at the ages of 3--4 Myr, cool winds do not contribute much to the energy injection rate. Thus, for both high- and low-$Z$ clusters, their energy injection rates do not rise drastically but continue to decline after 3--4 Myr, when the most massive stars end their lifetimes. 

We compare our mass-loss rates and energy injection rates with those adopted by the FIRE-3 cosmological simulations \citep{hopkins2022}, which are shown by the blue dash lines in Figure~\ref{fig:cluster_mdot}. The mass-loss rate in FIRE-3 is expressed by a function of time and $Z$ with four segments separated by 1.7, 4, and 20 Myr. 
For the high-$Z$ model, the post-main-sequence mass-loss rate from our simulations is in reasonable agreement with the FIRE-3 curve, while for the low-$Z$ model, our mass-loss rate after $\sim$4 Myr is about one order of magnitude lower than the FIRE-3 rate. The essential difference is because the low-$Z$ stars in our models do not evolve into cool supergiants when they enter the post-main-sequence stage, so the mass-loss rate is low. 
For the times before 4 Myr, the comparison of mass-loss rates is out of our scope as we do not consider the very massive stars with $>60\, M_{\odot}$. We also compare our energy injection rate the FIRE-3 function. For both high- and low-$Z$ cases, our energy injection rate is lower than the FIRE-3 rate by about a order of magnitude after $\sim$4 Myr.
FIRE-3 uses the entire mass-loss rate and the average velocity, which is $\sim 1,000$ km s$^{-1}$ at $\sim$4 Myr, to estimate the wind feedback. In contrast, we separately consider hot and cool winds, with hot winds having velocities of $> 1,000$ km s$^{-1}$ but low mass-loss rates and cool winds having high mass-loss rates but a velocity of 30 km s$^{-1}$. Thus, our calculations lead to a lower energy injection rate than the FIRE-3 function. 

We have calculated the feedback from stellar winds in a cluster using the Salpeter IMF. The critical metallicity of cool supergiant formation significantly affects the total mass loss within a cluster, and we suggest that this effect be considered in order to refine the treatment of stellar feedback in cosmological simulations. In contrast, the kinetic energy feedback is dominated by hot winds, which are irrelevant to the critical metallicity of cool supergiants.

\begin{figure*}[tbh]
\centering
\includegraphics[scale=0.5]{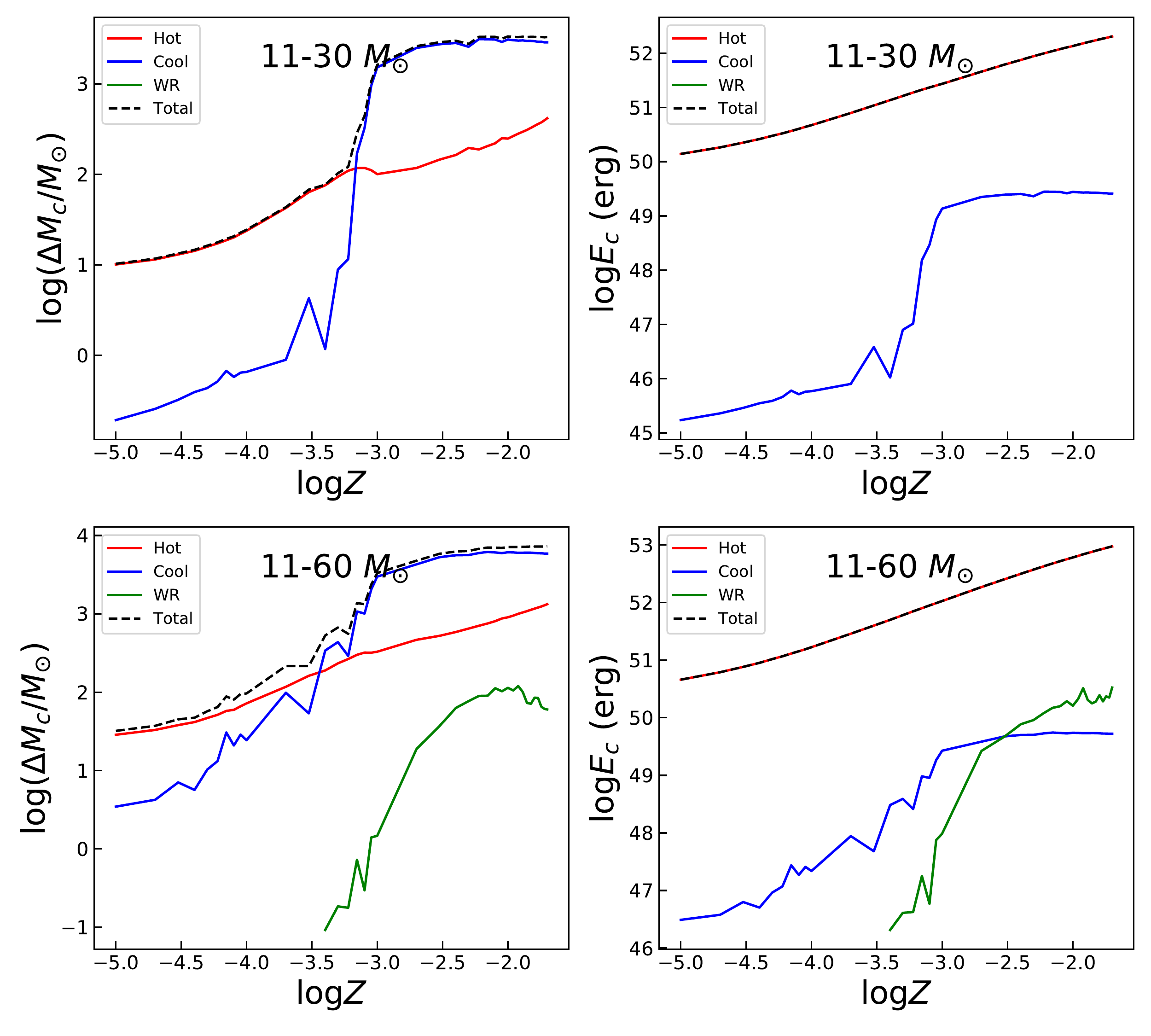}
\caption{The mass (left panels) and kinetic energy (right panels) released by the stellar winds in a star cluster of $10^5$ $M_{\odot}$, in which the Salpeter IMF is assumed. The upper panels only include the contribution from stars of 11--30 $M_{\odot}$, and the lower panels include the effect from all of our models within 11--60 $M_{\odot}$. The curves of mass loss show a significant increase around the critical metallicity of ($Z \sim 10^{-3}$), where the cool winds exceed the hot winds.  The corresponding kineitc energy curves are dominated by hot winds and thus do not show any sudden increase around the critical metallicity.}
\label{fig:feedback}
\end{figure*}

We further integrate the mass and energy injected by a $10^5$ $M_{\odot}$ cluster over time, and the results are shown in Figure~\ref{fig:feedback}. The contribution of wind feedback in our calculations is from star with $M_{\rm i} = 11-60\, M_{\odot}$. For comparison, we also plot the feedback that only includes the contribution of 11--30 $M_{\odot}$ stars. From these plots, hot winds dominate the kinetic energy and momentum feedback for all of the metallicities, while the cool winds surpasses hot winds in the contribution to mass loss when $Z>Z_{\rm{c}}$. The effect of critical metallicity occurs in the output of mass, but not in that of kinetic energy. 
The WR wind is never the dominating component, although its contribution to kinetic energy is larger than cool wind in an environment of solar metallicity. We note that these integrated quantities are the lower limit, as we do not consider the very massive stars with $>60\, M_{\odot}$.


\section{Discussions}\label{sec:discussions}

This section discusses some implications of and remaining issues with our results. In Section 6.1, we discuss the effects of metallicity on cool supergiant evolution. In Section 6.2, we point out the possible uncertainties in our models. Finally, in Section 6.3, we discuss the implications of critical metallicity on the fates and environments of massive stars.

\subsection{Two Effects of Metallicity on RSG Evolution}

Our study rinds that the critical metallicity of cool supergiant formation causes a significant increase in mass loss. This effect arises from stellar evolution rather than the treatment of mass-loss prescription. We discern two different effects of metallicity on RSG evolution and review the previous theoretical works that have shown a similar result of critical metallicity.

From the evolutionary tracks shown in Figure~\ref{fig:HR-Z}, there are two apparent effects of metallicity on the evolution of cool supergiants. First, at a lower metallicity, the overall evolutionary tracks, from the ZAMS to the supergiant stage, shift to higher $T_{\rm{eff}}$ or bluer locations. Second, the evolutionary tracks of stars with $Z<Z_{\rm{c}}$ end at hotter $T_{\rm{eff}}$, whereas stars with $Z>Z_{\rm{c}}$ can successfully evolve into cooler regimes. This is a sharp transition rather than a continuous change. In the following paragraphs we discuss these two different effects.

(1) \textit{The overall shift of the evolutionary tracks with metallicity.} This effect has been discussed with regard to the two ends of stellar evolution: the shift of ZAMS and the color of supergiants. First, the displacement of ZAMS toward the bluer locations with decreasing $Z$ has been reported by stellar models \citep[e.g.,][]{schaller1992,mowlavi1998,groh2019}. Among several physical parameters that directly change with metallicity (opacity, energy generation rate, molecular weight, etc.), the ZAMS shift has been attributed to opacity $\kappa$ \citep[see e.g.,][]{maeder2009}. In general, decreasing $Z$ leads to lower $\kappa$, transferring more radiation outward and leading to higher luminosities, and the stars also need to contract further to maintain the energetic equilibrium. With a larger $L$ and a smaller $R$, a higher $T_{\rm{eff}}\,(\propto L/R^2)$ is required for lower-$Z$ stars. 

At the end of stellar evolution, supergiants with lower $Z$ also shift toward bluer locations in the HR diagram. This feature has been observed in stars in the local group \citep {humphreys1979a,humphreys1979b,elias1985}, and has been reproduced in stellar models \citep[e.g.,][]{schaller1992,meynet1994,groh2019}.  Surveys of massive stars have confirmed that the RSGs in galaxies with lower metallicities have earlier spectral types on average: M2 I in the Milky Way, M1-M1.5 I in the LMC, and K3–K7 I in the Small Magellanic Cloud (SMC) \citep{massey2003,levesque2006}. 
This can be explained as the shift of the Hayashi line, which represents the coolest extent of giant stars, to higher $T_{\textrm{eff}}$ at lower $Z$ \citep{elias1985,levesque2006}. These stars shift to bluer positions at lower $Z$ due to opacity \citep{hayashi1961,kippenhahn2012}, which is a similar physical effect to the ZAMS shift.

(2) \textit{The critical metallicity of cool supergiant formation.} Instead of causing continuous shifts in the HR diagram, the critical metallicity acts as a boundary between two different types of evolution tracks in the supergiant stage. The stars with $Z>Z_{\rm{c}}$ successfully become RSG with $R \gtrsim 1,000\, R_{\odot}$, and those with $Z<Z_{\rm{c}}$ remain as smaller BSGs. This effect causes the significant increase in mass loss.

Some previous models have shown that low-$Z$ stars can never evolve into RSGs \citep{arnett1991,
baraffe1991,brocato1993,hirschi2007,eleid2009,limongi2017,groh2019}.
The metallicity value has been identified as $Z\sim 10^{-3}$\citep{baraffe1991,eleid2009}, which is consistent with our results. We start from the investigation of mass loss and identify the same phenomenon in cool supergiant formation. 

However, the physical origin of the critical metallicity of cool supergiant formation is still poorly understood. 
A profound question remains: Why do massive stars of $Z>Z_{\rm{c}}$ successfully expand to RSGs, but those of $Z<Z_{\rm{c}}$ fail to become RSGs? In companion Paper II, we will present some experiments that study the physical origin of critical metallicity. The explanation of its mechanism will help us understand the fundamental physics of RSG formation.

\subsection{Possible Uncertainties}

In this subsection, we discuss the uncertainties in our models from both stellar models and mass-loss prescriptions. We then evaluate the impacts of these factors on our main results.

\subsubsection{Uncertainties from stellar models}
For simplicity, this work only considers single and non-rotating stars. In fact, many recent studies of stellar evolution consider the effects of rotation and binarity \citep[e.g.,][for review]{langer2012}.

Stellar rotation can change the evolution tracks through processes such as rotational mixing \citep{maeder2009}. For RSGs, rotation is suggested to prevent a large intermediate convective zone, thus preventing the star from evolving backward to the blue side \citep{maeder2009}. It has also been suggested that at low metallicities, evolving into RSGs is easier for rotating stars than for non-rotating stars \citep{hirschi2007}.

Binarity is another important factor in recent stellar evolution studies, as most massive stars are found in binaries \citep{sana2012,duch2013}. Mass transfer is one of the essential processes in interacting binary stars that dramatically changes the post-main-sequence evolution \citep[e.g.,][]{iben1991,taam2000,yoon2010,langer2012,smith2014}. For example, an RSG can strip off its envelope due to mass transfer and move back to bluer locations in the HR diagram, becoming a YSG or WR star \citep{levesque2017}. 

Therefore, for rotating and binary stars, we cannot presume that the feature of critical metallicity is the same as our results of non-rotating and single stars. Furthermore, internal mixing processes such as semiconvection and overshooting can affect the evolution of supergiants \citep{schootemeijer2019}, and thus they may impact the critical metallicity.
The effects of rotation, binarity, and mixing processes on cool supergiant formation and the corresponding mass loss are thus worth studying in future works.

Another minor issue in our model is that some models of 11--17 $M_{\odot}$ do not successfully run until the iron core collapses. Instead, they stall at the carbon or oxygen burning stage. This situation does not affect our results, as the criterion of high or low mass loss refers to whether a star becomes a cool supergiant at the beginning of the core-helium burning stage. The final evolutionary stages only last for a very short time and thus do not contribute any significant mass loss via steady-state winds. Moreover, the models of $>$17 $M_{\odot}$, which smoothly evolve until the iron core collapses, already unambiguously demonstrate the feature of critical metallicity.

We further point out the limits of 1D stellar evolution models.  First, mixing in a 1D stellar model is done through the mixing length theory, which assumes all the chemical elements homogeneously mix together within the local pressure scale height if fluid instabilities occur. However, the mixing length theory likely becomes invalid inside the shell-burning region where violent burning, dynamics, and nucleosynthesis coevolve.  Precise mixing then becomes challenging.  Since a full 3D stellar evolution model is still unavailable, recent efforts \citep{tranpedach2013,tranpedach2014a,tranpedach2014b,jorgensen2018,mosumgaard2018}  have started to use local or global 3D hydro simulations of stellar convective zones to physically calibrate the parameters of mixing length theory used in 1D stellar evolution models. 

Another critical issue is the opacities of the stellar atmosphere.  One-dimensional stellar evolution models of MESA primarily focus on interior nuclear burning and structure evolution, and their stellar envelopes contain numerical artifacts due to their poor resolution in the Lagrangian code. In addition, the gas in the stellar atmosphere is at non-local thermal equilibrium, and its ionization states are difficult to calculate correctly. These issues can alter the surface opacities and significantly affect the physical properties of wind. Solving this problem requires multi-D radiation-hydro simulations, including sophisticated atomic physics, to evolve the wind using the first principles.

\subsubsection{Uncertainties from mass-loss prescriptions}

The currently available mass-loss prescriptions are based on the fitting of data obtained from simulations or observations, as described in Section \ref{sec:method}.  The best-effort prescriptions remain uncertain, and various prescriptions have been proposed. A comprehensive review of these prescriptions was given by \citet{mauron2011}.
More recently, some new prescriptions have been proposed for hot winds \citep[e.g.,][]{bjorklund2022} and cool winds \citep[e.g.,][]{goldman2017,beasor2020}. A new study of RSG mass loss by \citet{beasor2020} suggested that dJ88 overestimated the mass-loss rate by a factor of nine.

To explore the impact of uncertain mass-loss rates on stellar evolution, \citet{renzo2017} performed a grid of stellar evolution simulations with MESA using different wind prescriptions.
They also re-scaled the wind prescriptions by multiplying the mass-loss rates by a wind efficiency 
parameter $\eta$ and carried out simulations using $\eta$ of 0.1, 0.33, and 1. They found $\sim$50\% variations in the final stellar mass within all the combinations of $\eta$ and wind prescriptions and $\sim$15--30\% changes in the final stellar mass for a fixed $\eta=$1.

Another limit in our study is that we only consider steady-state winds. In addition to steady-state winds, the LBVs have eruptive mass loss driven by super-Eddington winds \citep{humphreys1994,shaviv2000,smith2006,owocki2017,vink2018,owocki2019}, such as $\eta$ Carinae.
These eruptive LBV winds may be a promising source of metal enrichment for Pop III stars in the early universe \citep{smith2006}.

The physical properties of a star cluster can also affect the mass loss of its resident stars. For example, in a cluster with multiple stellar populations, as a new star starts to form in the dense stellar environment shaped by the old stars, the accretion disk of its proto-star is disrupted by the gravity of old stars, leading to the formation of a rapidly rotating star that  evolves longer during its red giant phase, and produce more mass loss  \citep{tailo2015,tailo2020,tailo2021}. Although this environmental effect is based on the low-mass stars, we speculate that it may apply to forming massive stars and altering their mass loss.

The extension of the mass-loss prescriptions to low metallicities is another issue. For hot winds, \citet{vink2001} compared their mass-loss prescription with observations of stars in the SMC, where $Z\sim 1/5$ $Z_{\odot} $, and found reasonable agreement.
It has been suggested that at $Z\sim 1/7$ $Z_{\odot} $ mass-loss rates derived from non-LTE stellar atmospheric model fittings of the optical spectra cannot be described by the V01's prescription  \citep{tramper2011,tramper2014}. Nonetheless, subsequent analyses of far-UV spectra of the same targets indicate that \citet{tramper2011}'s low-Z stars actually have metallicities $\sim 1/5$ $Z_{\odot} $, making the mass-loss rates consistent with V01's prescription \citep{bouret2015}.
In regards to the cool winds, dJ88's prescription has also been tested in the SMC \citep{mauron2011}. Generally speaking, the mass-loss prescriptions have been tested in low-$Z$ stars with $Z\sim 1/5$ $Z_{\odot} $. For even lower metallicities, it is not known whether any breakdown of these scaling relations occurs. Few extremely metal-poor stars have been observed, and most of them are low-mass stars \citep{beers2005,yong2013,dacosta2019}, thus yielding no information on the mass-loss rates of massive stars. For now, adopting the widely-used mass-loss prescriptions in low-Z regimes is the only feasible method.

It is also unclear what kind of $Z$-dependent function should be included in mass-loss rate, as uncertainty exists, especially for cool winds. Earlier prescriptions of cool winds do not include the Z-dependence \citep{dejager1988,nieuwenhuijzen1990,vanloon2005}, but some later works adopt $\dot{M} \sim Z^m$ in their simulations \citep[e.g.,][]{mauron2011,groh2019}. Nevertheless, some recent observations show that mass-loss rates of RSGs are nearly independent of Z \citep{goldman2017,beasor2020}. We use the conventional dJ88 prescription without $Z$-dependence, but are aware of uncertainties in this function.

Fortunately, most of these uncertain factors in mass-loss prescriptions only slightly affect our main results. The major difference in mass loss between high- and low-$Z$ stars depends on whether a star evolves into the RSG phase rather than the mass-loss prescriptions. In other words, the physical origin of the critical metallicity lies in stellar evolution, regardless of the adopted recipe of mass loss, although these two processes can couple together to some extent. We carried out a simple test and found that even if the mass loss is fully turned off, the critical metallicity of cool supergiant formation still exists. As long as the mass-loss rate in the RSG phase is much higher than that in the main-sequence phase, the critical metallicity of cool supergiant formation leads to the bimodal distribution of mass loss. Therefore, whether the mass-loss rate is a function of $\dot{M} \sim Z^m$ is only a minor issue.

\subsection{Implications of the Critical Metallicity}

Our study reveals a critical metallicity of cool supergiant formation that leads to a significant difference in mass loss between high- and low-$Z$ environments. The low-$Z$ stars do not evolve into cool supergiants, and thus only have minimal mass loss through steady-state winds over their lifetimes. In the following paragraphs, we discuss the consequences of the critical metallicity.

It is widely known that pre-SN evolution and mass loss set up the physical condition for SN explosions. The types of progenitor stars often determine the types of SNe. For example, RSGs are the progenitors of SNe II-P, as identified in pre-SN images \citep[][for review]{smartt2009,smartt2015}. 
It has also been shown that the types of SNe can be significantly affected by the interaction with CSM produced by RSGs \citep[e.g.,][]{vanloon2010,ekstrom2012,smith2014,smartt2015,beasor2020,beasor2021,moriya2021}. 
For example, the early-phase light curves of SNe II-P can be better explained if dense CSM is considered \citep{moriya2011,moriya2017,moriya2018}. 
Moreover, it is suggested that CSM not only affects SN types but also makes SN explosions more likely to happen \citep{morozova2018}. 

Based on these understandings and the critical metallicity we found, we can expect the low-$Z$ stars to behave differently from high-$Z$ stars in SN explosions. First, if the low-$Z$ stars indeed end up as BSGs instead of RSGs, whether they will still explode as SNe is uncertain. Even if they successfully explode, the resulting SN type may differ from SNe II-P. Low-$Z$ stars never go through the RSG phase, or only become RSGs in the very final stage, so they only lose a negligible amount of mass and may not end as typical SNe II-P. 
Their circumstellar material can be too dilute for interacting SNe, although we have not considered eruptive mass loss that can generate dense CSM.

On a larger scale, the critical metallicity also impacts the stellar feedback in a star cluster. From our results, the total mass returned to the ISM by stellar winds shows a significant increase at $Z_{\rm{c}}$. In the high-$Z$ ($Z>Z_{\rm{c}}$) domain, these results are consistent with the current treatment of feedback in cosmological simulations such as FIRE-3. However, if the low-$Z$ ($Z<Z_{\rm{c}}$) stars do not evolve into cool supergiants, as our simulations predict, the current treatment of feedback may overestimate the mass injected to the ISM in these low-$Z$ environments. This discrepancy exists only in mass injection but not in energy output, as the kinetic energy feedback is dominated by hot winds, which are irrelevant to the critical metallicity.

As the critical metallicity $Z_{\rm{c}} \sim 0.001$ is lower than $0.1\, Z_{\odot}$, it is still difficult to test it using current observational instruments. Nevertheless, it will be promising to test the stellar evolution below this metallicity with the next-generation telescopes. For example, observations of massive stars in the low-$Z$ galaxies are listed as one of the scientific goals of the LUVOIR observatory \citep{garcia2019}.

In future works, we plan to apply these 1D mass loss models to 2D and 3D radiation hydro simulations to further understand how the mass loss shapes the CSM and ISM around massive stars.  With the advancement of numerical models and observational data, we may reveal the mystery of the mass loss of massive stars and their feedback.

\section{Conclusions}\label{sec:conclusions}

We studied the total mass loss of massive stars and their dependence on metallicity by performing 1D stellar evolution simulations with MESA. In the initial mass range 11--60 $M_{\odot}$, the total mass loss increases dramatically if their metallicity becomes higher than $Z\sim 10^{-3}$. We call this \textit{critical metallcity}, though the exact metallicity values for this mass-loss jump are not very well defined. Cool winds that operate in the cool supergiant phase are responsible for this mass-loss jump.  Massive stars with $Z<Z_{\rm{c}}$ usually stay blue during the post-main sequence; thus, the cool winds are not operating, and the mass loss is low. In contrast, massive stars with $Z>Z_{\rm{c}}$ successfully become cool supergiants, leading to significantly higher mass loss. 
The critical metallicity of cool supergiant formation gives rise to the bimodal distribution of mass loss.

We also calculated the feedback of stellar winds in a star cluster. In low-$Z$ environments, we may overestimate the integrated mass loss in a cluster if we do not consider the effect of critical metallicity. While the mass loss of a cool supergiant is much higher than that of a hot main-sequence star, its wind velocity is much lower, so its kinetic energy output is lower. Consequently, the critical metallicity affects mass loss but not kinetic energy feedback.

In summary, we identified the critical metallicity of cool supergiant formation in our stellar models. A striking consequence of the critical metallicity is the significant jump in mass loss when we adopt the widely used wind prescriptions. This evolutionary effect may have various consequences on the fates of metal-poor stars in the early universe.

\begin{acknowledgments}
This research is supported by the Ministry of Science and Technology, Taiwan under grant no. MOST 110-2112-M-001-068-MY3 and the Academia Sinica, Taiwan under a career development award under grant no. AS-CDA-111-M04.
\end{acknowledgments}


\end{CJK*}
\appendix
\section{Wind Prescriptions}\label{sec:appA}
This appendix presents the details of the mass-loss rate and wind velocity functions that we adopt in our simulations. 
\subsection{Mass-loss rate}
The settings of mass-loss rate in our simulations follow the wind schemes established in MESA. We apply the "Dutch" scheme for hot and WR winds and the dJ88 prescription for cool winds. The criteria to activate these winds are as follows:\\
(1) If $T_{\textrm{eff}}>12,000$ K, the "Dutch" scheme (\textit{hot/WR wind}) is fully operating. \\ (2) if $T_{\textrm{eff}}\le 8,000$ K, the \textit{cool wind} is fully operating, and the prescription of dJ88 is used. \\
(3) Between 80,000 K and 12,000 K, a linear interpolation of the dJ88 (cool) and "Dutch" (mainly hot/ WR) winds is applied. \\

The "Dutch" scheme is not a single function, but a combination of V01's hot wind prescription, NL00's WR wind prescription, and the extension of dJ88's cool wind prescription. In detail, the wind configuration of the "Dutch" scheme is as follows:\\
(1) If $T_{\textrm{eff}}\ge 11,000$ K and $X_H \ge 0.4$, the V01's hot wind prescription is adopted.\\
(2) If $T_{\textrm{eff}}\ge 11,000$ K and $X_H < 0.4$, the NL00's WR wind prescription is adopted. \\
(3) If $T_{\textrm{eff}}\leq 10,000$ K (low-T "Dutch" scheme), the dJ88 prescription, which is also used to express the cool wind, is adopted. \\
(4) If $10,000 \leq T_{\textrm{eff}}\leq 11,000$ K, a linear interpolation of dJ88 and V01/NL00 prescriptions is adopted.\\

There are criteria to turn off the mass loss at the very end of the stellar lifetime. If the central temperature is higher than $2\times 10^9$ K, mass loss will be fully turned off. If the central temperature is between $1 \times 10^9$ and $2 \times 10^9$ K, the mass-loss rate weakens linearly with the central temperature. These settings simply follow the default in MESA.

\subsection{Wind velocity}

As our MESA simulations do not include wind models, the wind terminal velocity ($v_{\infty}$) is estimated using empirical functions of $v_{\rm{esc}}$ and other stellar parameters. \\
\\
(1) \textit{Hot wind}. If the V01's prescription of mass loss is turned on, we follow the wind velocity formulae adopted by \citet{vink2001}. These expressions originate from \citet{leitherer1992} and \citet{lamers1995}.\\
a. If $T_{\textrm{eff}}>27,500$ K, which is higher than the bi-stability jump, then
\begin{equation}
v_{\infty}/v_{\rm{esc}}=2.6(Z/0.019)^{0.13}.
\end{equation}\\
b. If $T_{\textrm{eff}}<22,500$ K, which is lower than the bi-stability jump, then
\begin{equation}
v_{\infty}/v_{\rm{esc}}=1.3(Z/0.019)^{0.13}.
\end{equation}\\
c. If $22,500<T_{\textrm{eff}}<27,500$, the mass-loss rate is the linear combination of the high-$T$ mass-loss rate ($\dot{M}_h$) and the low-$T$ mass-loss rate ($\dot{M}_l$):
\begin{equation}
\dot{M} = \alpha \dot{M}_h + (1-\alpha) \dot{M}_l,
\end{equation}
where $\alpha \equiv (T_{\textrm{eff}}-22,500)/500$. The wind velocity we adopt is the weighted average of the high-$T$ wind velocity ($v_{\infty,h}$) and the low-$T$ wind velocity ($v_{\infty,l}$), and weighting coefficients are based on ratio of $\dot{M}$ from high- and low-$T$ components. The weighted average velocity $\bar{v}_{\infty}$ can be expressed as \\
\begin{equation}
\bar{v}_{\infty} = \frac{\alpha \dot{M}_h v_{\infty,h}+ (1-\alpha) \dot{M}_l v_{\infty,l}}{\alpha \dot{M}_h + (1-\alpha) \dot{M}_l}.
\end{equation}\\
(2) \textit{Cool wind}. We follow \citet{leitherer1992} to set the cool wind velocity a constant 30 km s$^{-1}$. As long as the dJ88 prescription is used, even if it is called from the "Dutch" low-$T$ scheme, we regard this part of mass loss as the cool wind. \\
\\
(3) \textit{WR wind.} We use the wind velocities given by \citet{nugis2000} for WR stars.\\
a. WN stars: 
\begin{equation}
\log (v_{\infty}/v_{\rm{esc}}) = 0.61-0.13 \log L + 0.3 \log Y.
\end{equation}\\
b. WC stars:
\begin{equation}
\log (v_{\infty}/v_{\rm{esc}}) = -2.37-0.43\log L -0.07 \log Z.
\end{equation}\\

In some transition temperatures, the mass-loss rate is composed of more than one wind component. In such cases, we use the ratios of mass-loss rates of these components as the weighting coefficients to calculate the average wind velocity. The expression is similar to Equation (A4), but with a linear combination of the hot/WR and cool wind velocities.

\end{document}